\documentclass[final,3p,times]{elsarticle}

\usepackage[utf8]{inputenc}

\title{An overview of experimental results from ultra-relativistic heavy-ion collisions at the CERN LHC: bulk properties and dynamical evolution}

\usepackage{graphicx}
\usepackage{amssymb}
\usepackage{amsmath}
\usepackage{bm}
\usepackage{color}
\usepackage{float}
\usepackage{caption}
\biboptions{sort&compress}
\usepackage{xspace}
\usepackage{tabularx}
\usepackage{multirow}
\newcommand{\Pt}{$p_{\rm{T}}$}
\newcommand{\pt}{$p_{\rm{T}}$}
\newcommand{\pT}{$p_{\rm{T}}$}

\newcommand{\Jpsi}{J/$\psi$}
\newcommand{\jpsi}{J/$\psi$}

\newcommand{\pp}{pp}
\newcommand{\pA}{pA}
\newcommand{\pPb}{p--Pb}
\newcommand{\PbPb}{Pb--Pb}
\newcommand{\AuAu}{Au--Au}
\newcommand{\dAu}{d--Au}
\newcommand{\sqrsNNnoEq}{$\sqrt{s_{\rm NN}}$}
\newcommand{\sqrsNN}{$\sqrt{s_{\rm NN}}=$}

\newcommand{\vtwo}{$v_2$}

\newcommand{\Npart}{$N_{\rm part}$}
\newcommand{\Ncoll}{$N_{\rm coll}$}
\newcommand{\RAA}{$R_{\rm AA}$}

\newcommand{\RpA}{$R_{\rm pA}$}
\newcommand{\RpPb}{$R_{\rm pPb}$}

\newcommand{\W} {\ensuremath{W}}
\newcommand{\Z} {\ensuremath{Z}}

\newcommand{\snn}{\ensuremath{\sqrt{s_{\mathrm{\scriptscriptstyle NN}}}}\xspace}

\newcommand{\chic}{\ensuremath{\chi_c}\xspace}
\newcommand{\chib}{\ensuremath{\chi_b}\xspace}
\newcommand{\psiP}{\ensuremath{\psi\text{(2S)}}\xspace}

\newcommand{\upsa}{\ensuremath{\Upsilon\text{(1S)}}\xspace}
\newcommand{\upsb}{\ensuremath{\Upsilon\text{(2S)}}\xspace}
\newcommand{\upsc}{\ensuremath{\Upsilon\text{(3S)}}\xspace}

\newcommand{\upsn}{\ensuremath{\Upsilon\text{(nS)}}\xspace}

\newcommand{\Dmeson}[1]{\ensuremath{\mathrm{D}^{#1}}\xspace}

\newcommand{\Dzero}{\Dmeson{0}}
\newcommand{\Dplus}{\Dmeson{+}}

\newcommand{\Dstarplus}{\Dmeson{*+}}

\newcommand{\hfe}{\ensuremath{{\rm HF} \to e^{\pm}}\xspace}
\newcommand{\hfm}{\ensuremath{{\rm HF} \to \mu^{\pm}}\xspace}

\newcommand{\raa}{\ensuremath{R_{\mathrm{AA}}}\xspace}

\newcommand{\rcp}{\ensuremath{R_{\mathrm{CP}}}\xspace}

\newcommand{\TeV}{\ensuremath{\text{~TeV}}\xspace}

\newcommand{\GeVc}{\ensuremath{\text{~GeV}/c}\xspace}

\newcommand{\beq}{\begin{equation}}
\newcommand{\eeq}{\end{equation}}
\newcommand{\beqn}{\begin{eqnarray}}
\newcommand{\eeqn}{\end{eqnarray}}

\newcommand{\pb}{{Pb--Pb}\xspace}

\usepackage{lineno}
\begin{document}


\begin{frontmatter}

\title{ An overview of experimental results from ultra-relativistic heavy-ion collisions at the CERN LHC: hard probes}

\author[YF]{Panagiota Foka}
\ead{yiota.foka@cern.ch}
\author[MAJ]{Ma{\l}gorzata Anna Janik\corref{CorrespondingAuthor}}
\ead{majanik@if.pw.edu.pl}

\address[YF]{GSI Helmholtzzentrum f\"ur Schwerionenforschung GmbH, Planckstra\ss e 1, 64291 Darmstadt, Germany}
\address[MAJ]{Faculty of Physics, Warsaw University of Technology, Koszykowa 75, 00710 Warsaw, Poland}

\cortext[CorrespondingAuthor]{Corresponding author}

\begin{abstract}
The first collisions of lead nuclei, delivered by the CERN Large Hadron Collider (LHC) at the end of 2010, at a centre-of-mass
energy per nucleon pair \sqrsNN\ 2.76 TeV, marked the beginning of a new era in ultra-relativistic heavy-ion physics.
The study of the properties of the produced hot and dense strongly-interacting matter at these unprecedented energies is
currently experimentally pursued by all four big LHC experiments, ALICE, ATLAS, CMS, and LHCb.
The more than a factor 10 increase of collision energy at LHC, relative to the previously achieved maximal energy at other collider facilities, results in an increase of production rates of hard probes.
This review presents selected experimental results focusing on observables probing hard processes in heavy-ion collisions delivered during the first three years of the LHC operation.
It also presents the first results from Run 2 heavy-ion data at the highest energy,
as well as from the studies of the reference pp and \pPb\ systems, which are an integral part of the heavy-ion programme.

\end{abstract}


\date{\today}

\end{frontmatter}

\section{Introduction}
\label{sec:Introduction}
The aim of ultra-relativistic heavy-ion physics 
is to study strongly interacting matter under extreme conditions 
of high temperature and energy density,
where quantum chromodynamics (QCD), the theory of strong interactions within the Standard Model,
predicts  
a transition to
a new phase of matter, the quark-gluon plasma, QGP (i.e. see \cite{CasalderreySolana:2011us} and references therein).
The QGP is considered to be the QCD ground state,
where partons (quarks and gluons) are deconfined, i.e. no longer bound into composite particles. In addition, chiral symmetry is (approximately) restored, i.e. 
light quarks are (approximately) massless.
Such a state of matter existed in the primordial universe, microseconds after the Big Bang,
and may still exist today in the cores of neutron starts.

Based on the QCD calcularions on the lattice, the transition from normal (nuclear or hadronic) matter to the QGP is expected to occur at a critical temperature\footnote{In fact, it is a pseudo-critical temperature as 'lattice QCD' calculations indicate a crossover rather than a well defined phase transition \cite{Aoki:2006we,PseudoCritical}.} 
of the order of $\sim$200 MeV (more than $10^{12}$ K) -- the order of the QCD scale parameter, $\Lambda_{\rm QCD}$.
In order to achieve the conditions necessary for the formation of the QGP, 
a large volume of hot and dense matter is thought to be required, 
and therefore such research has been pursued with collisions of heavy nuclei at the highest possible collision energies.
Because the strong coupling constant at the energy scale of the processes relevant to the production of the bulk of the matter (i.e. soft sector) is large,
techniques such as pQCD are no longer applicable.
Therefore, the heavy-ion research field presents a unique opportunity, as well as a testing ground, of novel approaches to QCD in a new regime where the strong interaction is indeed strong.
Particularly at the high energy regime of Large Hadron Collider (LHC),
ultra-relativistic heavy-ion physics connects the better-known ``elementary-interaction" aspects of high-energy physics
with the ``macroscopic-matter" aspects of nuclear physics still to be explored.
Hence,
a novel, interdisciplinary approach to investigate matter along with its interactions is being developed
applying ideas and methods from both high energy and nuclear physics. 
Those span today from  computationally intensive numerical solutions (lattice QCD),
thermodynamical and statistical methods,
classical solutions in the high-density limit (Colour Glass Condensate)
up to quantum gravity (Conformal Field Theory in Anti-de-Sitter Space or AdS/CFT).

In general, such studies 
are expected to provide information on the properties of large, complex systems including elementary quantum fields, and an indication on 
the influence of the microscopic laws of physics, expressed by the ``QCD equations", on the macroscopic phenomena like phase transitions and critical behaviour.
In this context the study of nuclear matter and its different phases is of relevance also beyond the QCD specific domain, because phase transitions and symmetry breaking are principal concepts of the Standard Model and the QCD phase transitions are the only ones that are within reach of laboratory experiments.
In summary, the tasks of the heavy-ion research field is to search for the predicted QGP,
measure its properties, study and potentially discover QCD aspects in the non-perturbative sector.

Experimentally, this new and rapidly evolving research field
has already presented a wealth of experimental results
since the first pioneering experiments, started at relativistic energies in late 70s.
With the first ion beams at LHC the energy in the center-of-mass system per nucleon pair, \sqrsNNnoEq\ , increased
by four orders of magnitude in slightly more than 25 years.
At the time of the LHC startup, 
after about ten years of research at RHIC at \sqrsNNnoEq\ up to 200 GeV
and a similar time at fixed-target machines at about one tenth of this energy,
discovery of QGP is well established and the systematic characterization of its properties well advanced, a claim based on theoretical interpretation of a large sample of comprehensive experimental data available already before the startup of LHC \cite{Carminati:2004fp,Alessandro:2006yt,Adams:2005dq,Adcox:2004mh,Back:2004je,Arsene:2004fa}.
An overview of recent LHC results focusing on bulk, so-called soft observables can be found in the accompanying article~\cite{P1} published in the same journal and summarized below.

In contrast to the expectations that the QGP would have properties similar to an almost ideal, weakly coupled gas of quarks and gluons,
experimental results from RHIC, summarized in 2005 \cite{Adcox:2004mh,Arsene:2004fa,Back:2004je,Adams:2005dq},
have shown that a hot, strongly interacting, nearly perfect and almost opaque liquid, 
also called the sQGP (s standing for strongly interacting)
was produced in central (head-on) \AuAu\ collisions at the top RHIC energy.
The created medium has very small shear viscosity (therefore, it is characterized by very limited internal friction)
and responds to pressure gradients by flowing roughly unobstructed \cite{Muller:2006ee,Jacobs:2004qv}. 
Moreover, it is almost opaque -- most of the energy of fast partons propagating through it is absorbed.
Describing QGP as a ``fluid" indicates properties of ``macroscopic matter" and collective degrees of freedom
(within the hydrodynamic models framework), existing for a time significantly larger than the relevant relaxation times and with dimensions substantially larger than the mean free path.
The aspect of ``perfect liquid" was justified
from measurements of 
collective particle motion, known as``elliptic flow",
which develops as a response to the initial geometric conditions (reflected by the impact of the collision\footnote{For a detailed description of the collision geometry see i.e. \cite{CasalderreySolana:2011us}.}) and pressure gradients in the collision overlap region where the QGP is created, for details see \cite{P1}. 
The magnitude of elliptic flow at RHIC was found to exceed the maximum possible value
predicted by hydrodynamics for a given initial deformation, 
corresponding to the reaction of a perfect liquid with minimal shear viscosity over entropy density ratio,
which is reached in an extremely strongly interacting system with mean free path approaching 
the smallest possible value (the Compton wavelength).
The measurement of direct\footnote{Direct photons are photons not originating from hadron decays. They may originate from different stages of the collision, i.e. direct prompt photons coming from the initial hard parton scatterings, direct thermal photons originating from the QGP state.} ``thermal" photons radiated by the deconfined quark-gluon matter and its interpretation within hydrodynamic based models gave an estimate of the initial temperature of the hot liquid of at least about 300 MeV.
The opaque aspect of this liquid came from the observed suppression of high-\pt\ particles  (typically leading jet fragments)
relative to \pp\ collisions, by a factor of about 5
which is also an indication of very strong final state interactions.
It was verified with essential control measurements  that the suppression was
not seen in \dAu\ interactions (eliminating as the reason effects present in cold nuclear matter) 
as well as not seen with colour neutral probes, 
establishing, therefore, that the observed suppression in nuclear collisions  
is due to the strong interactions in the final state 
caused by the QGP. 

From the RHIC results, in less than 10 years, 
a ``Heavy-Ion Standard Model" (HISM) emerged, describing the dynamic evolution and characterizing the high density state created in 
ultra-relativistic heavy-ion collisions \cite{Heinz:2013wva,Schukraft:2011np}.
The current understanding is that the fireball created in such collisions is in local thermodynamic equilibrium well described by hydrodynamics, i.e. particle chemistry is in agreement with thermal model predictions and particle spectra show patterns of radial and elliptic hydrodynamic flow. 
Therefore, one of the main goals at LHC was to measure, with increased precision and new, unique probes, the parameters that characterize this new state of matter.
After verifying first that the global event characteristics, reflecting the bulk matter properties 
(such as energy density, volume, lifetime)
are indeed different at the LHC energy regime, 
while the evolution and the intrinsic medium properties are still
properly described by the HISM,
the LHC programme focused on precision measurements of the QGP parameters 
(i.e. equation-of-state, viscosity, transport coefficients, Debye screening mass).

First results at LHC came fast covering a variety of topics and painting the general picture while detailed multi-differential measurements are still being pursued.
The HISM could be probed, for the first time, in a higher energy regime \cite{Heinz:2013wva,Schukraft:2011np}; it was found to be robust enough and provided reliable extrapolations and predictions at both \PbPb\ energies (\sqrsNN\ 2.76 and 5.02 TeV) that LHC delivered so far.
Indeed, as expected 
at the higher center-of-mass energy of LHC the created matter was found to be characterized 
by larger energy density, freeze-out volume, and lifetime in comparison to RHIC,
while the most critical tests of HISM came from the experimental measurements of flow (azimuthal angle anisotrophy) observables at LHC which are found in agreement with HISM predictions \cite{Adam:2016izf,Adam:2016nfo}.
Detailed studies  
for a more precise determination of the shear viscosity
as well as its temperature dependence
studying \PbPb\ collisions at 2.76 TeV and first results at 5.02 TeV are presented in the accompanying article \cite{P1} together with further results exploring still unanswered questions of effects such as the hadronization phase and the interplay of soft and hard processes. 

The energy advantage of LHC is more apparent in the area of parton energy loss associated to the opaque nature of the sQGP
where the kinematic regime exceeds by far the one reached at RHIC.
The most important impact of the increase of the collision energy is the large increase of the rates of hard probes, such as jets, electro-weak particles and heavy flavours, including the full family of quarkonia ($c\bar{c}$ and $b\bar{b}$ bound states).
The available high rates make possible precision studies of the QGP using the interactions of these probes  with the medium constituents, which are under better theoretical control than the propagation of light partons \cite{Andronic:2015wma}. In addition, some observables, e.g. very high-energy jets, electro-weak bosons, and different $\Upsilon$ states, are accessible in heavy-ion collisions for the first time. 
A new generation of powerful, large-acceptance, state-of-the-art experiments, ALICE, ATLAS, CMS, and LHCb provided a great advantage that made this task possible.

This article presents a subjective selection of representative results of heavy-ion research from the first three years of LHC Run 1 with emphasis on the hard observables, 
abundantly available at LHC,
used to probe the created system. 
Together with the \PbPb\ results at \sqrsNN\ 2.76 TeV we also discuss results from pp and \pPb\ collisions 
most relevant to the hard probes and published at the time of writing. 
We also present in the same journal a similar review of results on global bulk matter properties and the dynamics of the created system 
accessible via soft probes \cite{P1}.
Other reviews of heavy-ion LHC results and references to the literature can be found in \cite{Andronic:2014zha,Brambilla:2014jmp,Armesto:2015ioy,Roland:2014jsa,Loizides:2016tew,Schukraft:2013wba, Muller:2012zq,Norbeck:2014loa,Prino:2016cni}.

\section{Energy loss}
\label{sec:heavy_intro}
The high energies reached in heavy-ion collisions at the LHC allow precision studies of hard processes that involve high momentum or mass scales, larger than any scale of the 
QGP  medium produced in the collision. 
Such probes originate from hard partonic scatterings at the very initial stage of the collision 
($\tau \sim 1/Q$, where $Q$ is the virtuality transfer), 
before the QGP is created, 
and therefore experience the full evolution of the created fireball. 
They are regarded for this reason as ``external probes". 
Hard probes can be computed in perturbative QCD; their production and propagation through the medium can thus provide the means to probe experimentally the nature and properties of the medium they traverse, through their interactions with its constituents.

One of the most studied medium induced effects is
the attenuation of jet yields 
(or modification of jet spectra)
due to the energy loss of the parent parton, commonly known as ``jet quenching", initially proposed by Bjorken in 1982 \cite{Bjorken:1982qr}.
Since then, theoretical advances  have established the studies of energy loss as a precision tool to probe the nature and properties of the traversed medium \cite{Wiedemann:2009sh}.

In a perturbative QCD framework energy loss is expected to occur  both via inelastic (radiative energy loss, via medium induced gluon radiation) \cite{Gyulassy:1990ye} and elastic (collisional energy loss) \cite{Thoma:1990fm} processes. 
Radiative energy loss dominates at high energies while elastic energy loss is expected 
to contribute at lower energies.

The amount of energy lost, $\Delta E$, is predicted to depend on the properties of the medium, in particular its opacity (associated to the medium density and the interaction strength) and the 
path length traversed inside the medium.
In general, the strength of the interaction of partons with the constituents of the medium is  characterized by the transport coefficients (for radiational energy loss usually given by $\hat{q}$,  
the average transverse momentum squared acquired by the parton per unit path length). 
Overall,
such studies probe different aspects of the energy loss mechanism, the interaction strength and properties of the medium as detailed \cite{Baier:1999ds,Salgado:2003rv}. 

The parton energy loss via medium-induced gluon radiation is predicted to decrease with increasing parton mass \cite{Dokshitzer:2001zm}.
The main reason for this is the so-called ``dead-cone effect",
introduced first for the vacuum radiation \cite{0954-3899-17-10-003} 
and then applied, in a similar way, to the medium-induced radiation 
\cite{Dokshitzer:2001zm}. 
Due to a destructive interference, the radiation is suppressed in directions close to that of the quark.
The heavier the quark is, the larger the exclusion region. 
In addition, at the LHC energies, the light flavoured-hadrons (at \Pt\ of order 10 GeV/$c$) 
mostly originate from gluon fragmentation \cite{Djordjevic:2013pba}
while heavy-flavoured hadrons are produced via the fragmentation of the corresponding heavy quarks. 
Because gluons have larger colour charge than quarks, and therefore larger colour coupling, they are expected to suffer more radiative energy loss in the deconfined medium.
Therefore, the expectation is that heavy quarks (charm and bottom) lose less energy 
compared to lighter ones (up, down, strange) \cite{Dokshitzer:2001zm}, 
leading to a hierarchy for the energy loss,  
$\Delta E_{\rm gluon} > \Delta E_{\rm light\ quark}>\Delta E_{\rm charm} > \Delta E_{\rm bottom}$ 
in the kinematic regime where the mass cannot be neglected with respect to the parton momentum.

The effect of energy loss is usually quantified through the nuclear modification factor, 
which is the yield of a given observable (such as charged hadrons, identified particles and/or reconstructed jets) measured in nucleus-nucleus collisions, AA,
properly normalized to the pp measurement at the same nucleon-nucleon energy:
\begin{equation}
\label{eq:RAA}
{R_{AA}(p_T)=\frac{{\rm d}N^{AA}(p_T)/{\rm d}p_T}{\langle N_{coll} \rangle {\rm d}N^{pp}(p_T)/{\rm d}p_T}}.
\end{equation}
If an AA collision behaved like a simple superposition of independent \Ncoll \footnote{In the framework of a geometrical model of heavy-ion collisions,  so-called Glauber model \cite{Miller:2007ri}, \Ncoll\ is defined as the number of single nucleon-nucleon collisions. From the same model  \Npart\ can be estimated,
which is the number of single nucleon-nucleon collisions.} \pp\ collisions, the \RAA\ would be equal \mbox{to 1}. 
However, for soft processes, such as particle production
at \pT\ below a few GeV, the scaling from pp to AA is governed by \Npart\ rather than by \Ncoll , leading naturally to an \RAA\ below
unity in that \pT\ region.
Departure of \RAA\ from unity signals a change of physics in AA collisions and provides input to quantify medium induced effects.
On the basis of the above arguments the effects of the QGP medium formed in the collision would  lead to an experimentally observed suppression pattern  $R_{AA}^{light}<R_{AA}^{\rm charm}<R_{AA}^{\rm beauty}$ \cite{Andronic:2015wma}.
%

The study of differential observables is expected to shed light into the different interaction mechanisms.
In particular, the dependence of the parton energy loss on the path length traversed in the medium 
is predicted to be linear for collisional energy loss \cite{Thoma:1990fm,Braaten:1991jj,Braaten:1991we} 
and close to quadratic for radiative processes in a QGP \cite{Baier:1996sk}
(and even a cubic dependence on the path length is predicted within the AdS/CFT framework).

%
%
The path-length dependence of energy loss can be probed experimentally by studying the dependence of the yields of hard probes on the collision centrality. 
Because the energy loss suffered by a parton depends on the traversed path length in the medium, 
it is expected that in central heavy-ion collisions the total energy loss will be larger than in peripheral ones.
However, when comparing results of central and peripheral collisions one has to bear in mind the differences of the medium conditions (central collisions are expected to create a hotter, denser medium).  
%
%
Further insight into the path-length dependence of parton energy loss is expected to be gained studying the dependence of the yields of hard probes on the azimuthal angle relative to the reaction plane in non-central \PbPb\ collisions \cite{Aad:2013sla}.
Non-central heavy-ion collisions create an initially almond-shaped collision zone where pressure gradients are developed due to reinteractions of the created medium, and transform the initial spatial anisotropy to a measurable momentum anisotropy reflected to an azimuthal angle anisotropy. 
Such an anisotropy can be quantified by the so-called elliptic flow \vtwo\ \cite{P1}, which can be extracted from 
$v_{\rm 2} = \langle \cos \big [2(\varphi - \Psi_{2}) \big] \rangle \,$,
where $\Psi_{\rm 2}$ is the azimuthal angle of the 2$^{\rm nd}$ order symmetry plane of the overlap region, and $\varphi$ is the particle's azimuthal angle (by applying a Fourier decomposition to the measured distribution~\cite{Voloshin:1994mz}) as detailed in the accompanying article on soft observables \cite{P1}.
A larger suppression is expected out-of-plane than in-plane as partons traverse a larger path along the longer axis of the initially almond-shaped collision zone. 
Such studies allow the measurement of the relative energy loss that hard probes suffer traversing different lengths of the medium under the same medium conditions.

However, to be able to quantify any changes caused by the presence of the medium, it is important to 
establish the initial flux and  conditions with precise measurements of the total cross sections. It is also crucial to  compare the AA results systematically with reference \pp\ (and pA) collisions at the same centre-of-mass energy and an appropriate kinematic regime.

%
Experimentally, first manifestations of parton energy loss were observed  at RHIC,   
establishing that a very opaque partonic medium  
was created in  \AuAu\  central collisions at \sqrsNN\ 200 GeV \cite{Adcox:2001jp,Adler:2002xw,Adler:2002tq}.
Sophisticated, large acceptance, state-of-the-art detectors combining precision tracking and vertexing with calorimetry allow studies not only of high-\pT\ hadrons but also of reconstructed jets, and high-precision measurements of heavy-flavoured particles as well as of flavour-tagged jets. 

An added advantage at LHC is the availability of  probes that do not interact strongly with the medium such as real and virtual photons at high \Pt\ as well as electro-weak bosons. 
Such control probes, including the $W$ and $Z$ (decaying in the leptonic channels), reconstructed in heavy-ion collisions for first time, allow testing perturbative QCD in nuclear collisions.

The goal of systematic multi-differential studies of hard probes in pp, pA and AA collisions is to
shed light on the details of parton energy loss, 
disentangle the interplay of different mechanisms,
and ultimately  characterize the properties of the created medium.
%
Measurements of ``jet quenching" at LHC through the study of high-\Pt\ particles are presented in 
Sec.~\ref{sec:ExpResultsHeavy_HighPt} and results 
using reconstructed jets 
are discussed in Sec.~\ref{sec:ExpResultsHeavy_Jets}.  
Further details on heavy-flavour energy loss as well as the expected flavour and path-length dependence of parton energy loss are described in Sec.~\ref{sec:ExpResultsHeavy_HeavyFlav}
while Sec.~\ref{sec:ExpResultsHeavy_Quarkonium} is focusing on properties of quarkonia bound states and their modification due to the presence of the medium.

\section{High \pt\ particles}
\label{sec:ExpResultsHeavy_HighPt}

\begin{figure}[hbt]

    \includegraphics[width=0.45\textwidth]{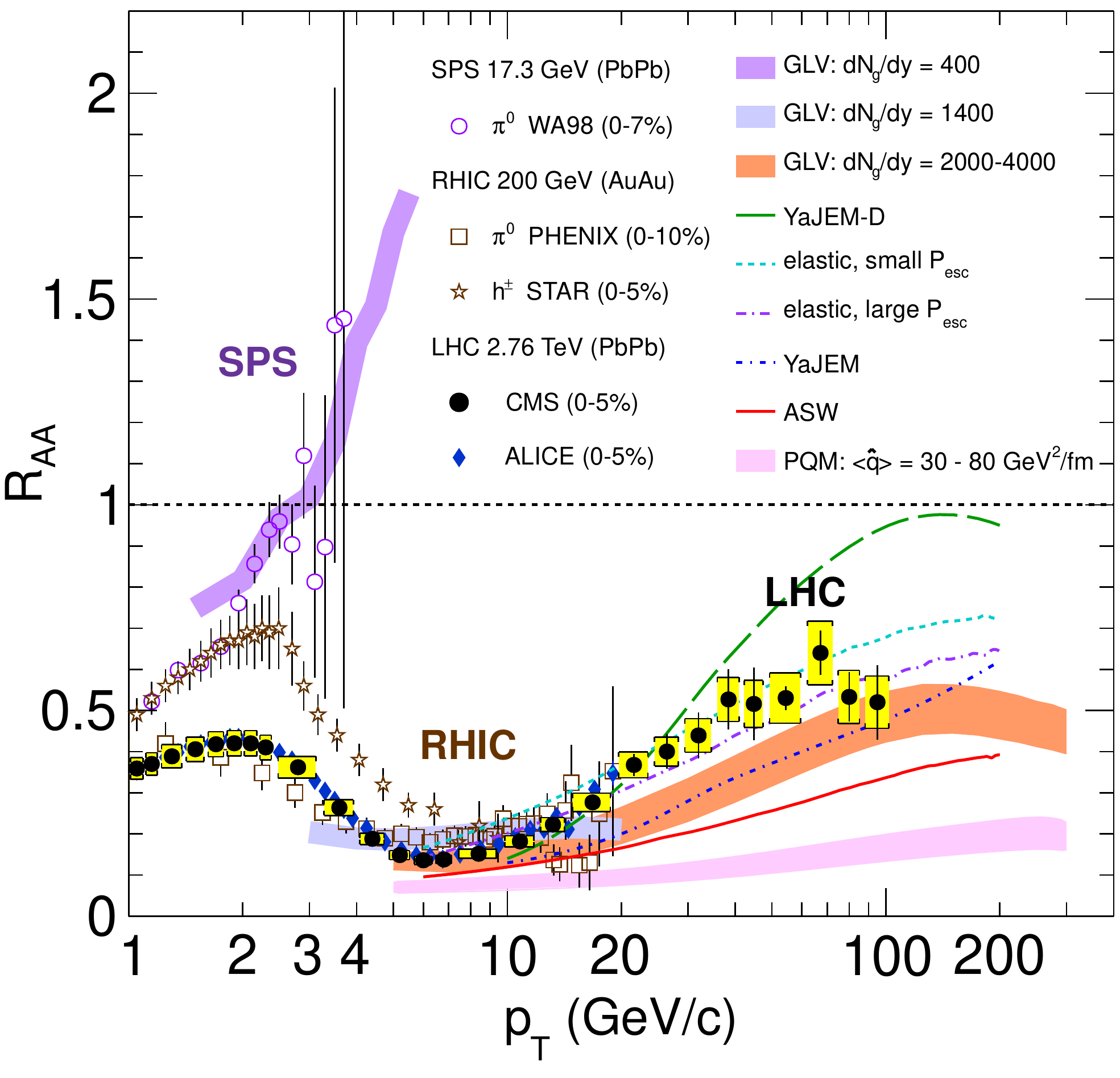}  
    \includegraphics[width=.55\textwidth]{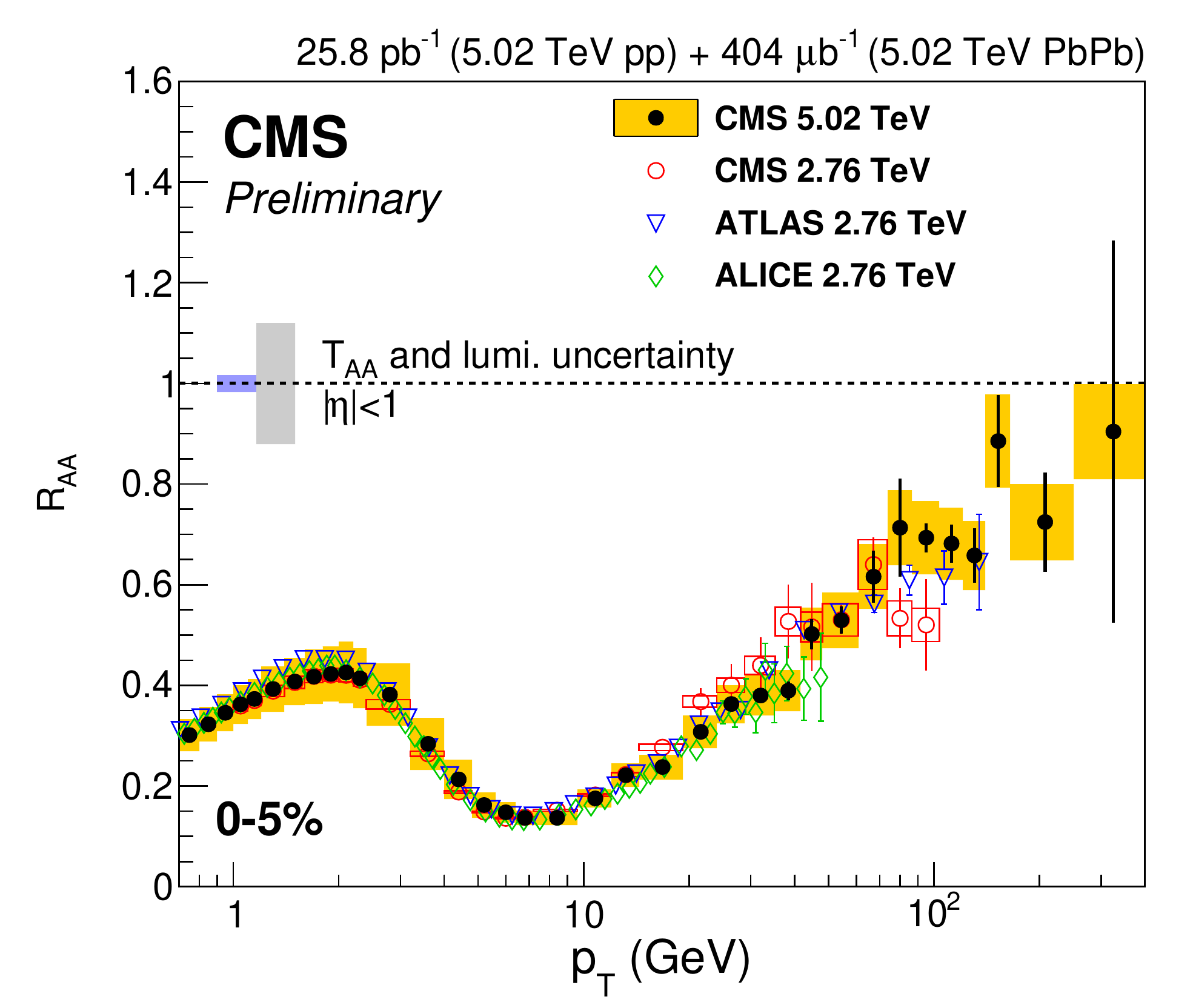}  
    \caption{
        (Left)   
        Measurements of the nuclear modification factor RAA in central heavy-ion collisions
at three different $\sqrt{s_{\rm NN}}$, as a function of \pT, for neutral pions ($\pi^0$),
charged hadrons ($h^{\pm}$), and charged particles \cite{CMS:2012aa,Aamodt:2010jd, Aggarwal:2001gn, d'Enterria:2004ig, Adare:2008qa, Adams:2003kv}, compared to several theoretical predictions (for references see \cite{CMS:2012aa}). Figure from \cite{CMS:2012aa}.  
    (Right)         
     Charged particle \RAA\ measured by CMS in 0--5\% centrality interval at \sqrsNN\ 5.02 TeV \cite{CMS-PAS-HIN-15-015} compared to CMS \cite{CMS:2012aa}, ALICE \cite{Abelev:2012hxa} and ATLAS \cite{Aad:2015wga}  results at \sqrsNN\ 2.76 TeV.
    Figure from \cite{CMS-PAS-HIN-15-015}. 
         }
    \label{fig:RAA2} 
\end{figure}

The nuclear modification factor of charged particles  for SPS, RHIC and LHC is shown in Fig.~\ref{fig:RAA2}-left \cite{CMS:2012aa,Aamodt:2010jd, Aggarwal:2001gn, d'Enterria:2004ig, Adare:2008qa, Adams:2003kv}.
The LHC measurements show a slightly stronger suppression than those from RHIC \cite{Adcox:2001jp,Adler:2002xw}; the largest measured suppression, in the \pT\ range 6--9 GeV/$c$, is at the LHC a factor of about 7, 
while at RHIC a factor of 5 was observed. 
A completely new observation at the LHC is that with increasing \pT\ the suppression becomes smaller, i.e. \RAA\ increases. 
By extending the \pT\ up to 300 GeV/$c$ for \sqrsNN 5.02 TeV, see Fig.~\ref{fig:RAA2}-right, CMS \cite{CMS:2012aa,CMS-PAS-HIN-15-015}
showed that the 
maximal suppression is followed by a rising trend up to the highest transverse momenta measured in the analysis. 
This demonstrates that even very energetic partons of the highest \Pt\ suffer considerable energy loss interacting with the medium. 

\begin{figure}[hbt]
    \includegraphics[width=.55\textwidth]{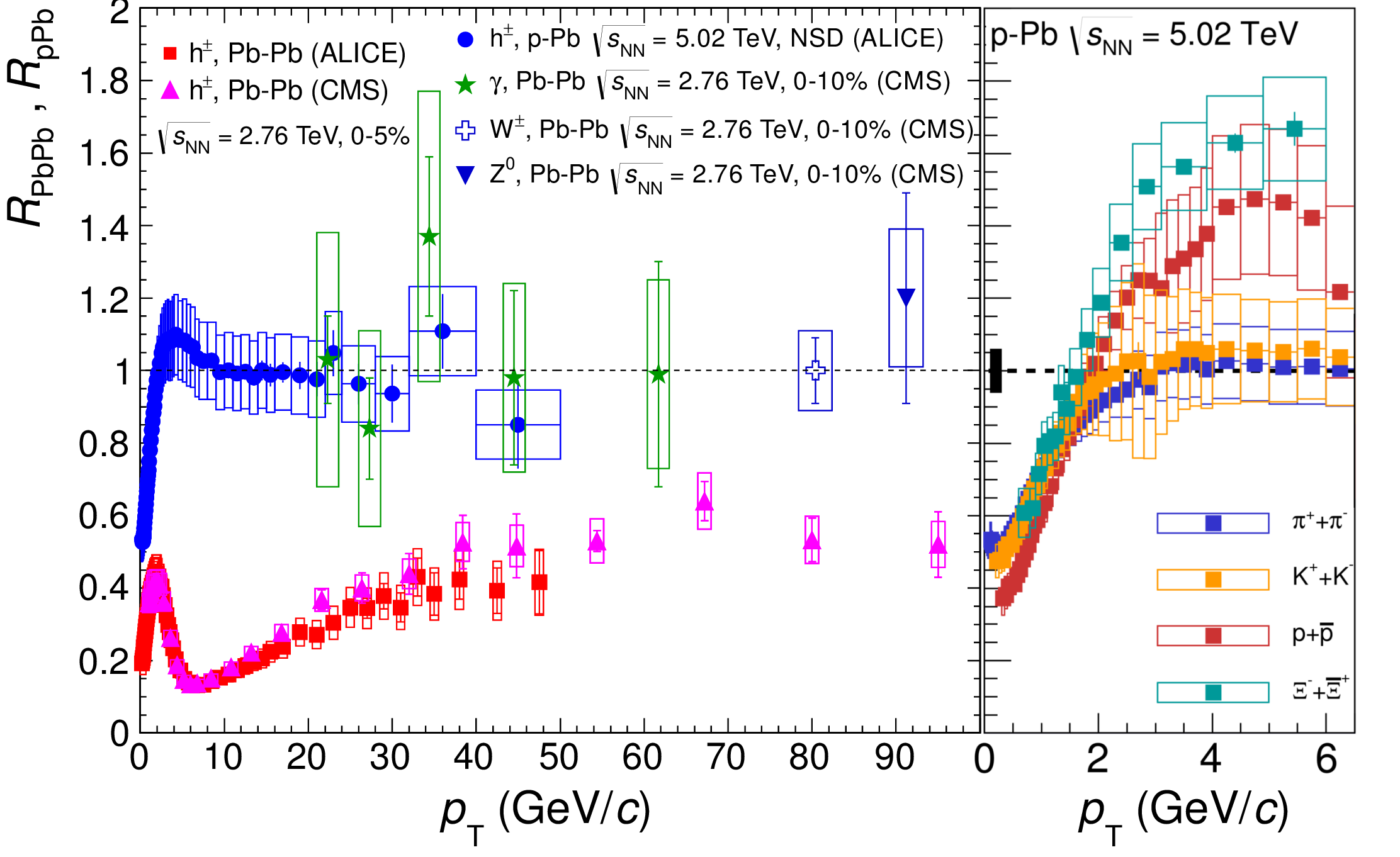}
    \includegraphics[width=0.475\textwidth]{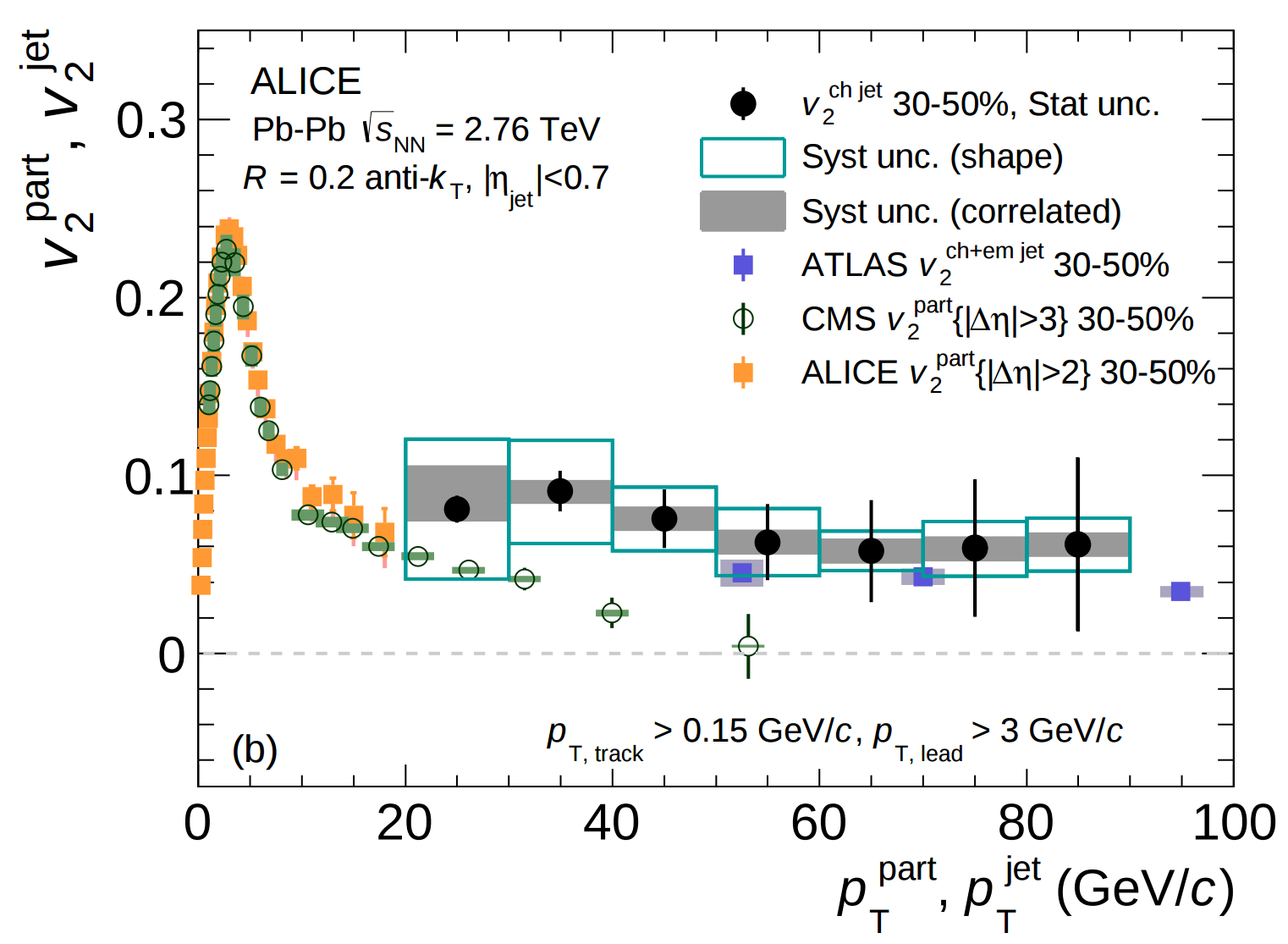}    
    \caption{(Left) The \pt\ dependence of the nuclear modification factor \RpPb\ in central 
            Pb--Pb collisions at \sqrsNN\ 2.76 TeV
            compared to the nuclear modification 
            factor \RpPb\ of charged particles (h$^\pm$)  in
            \pPb\ collisions at \sqrsNN\ 5.02~TeV \cite{Abelev:2014dsa}.
        The \PbPb\ data are for charged particle \cite{CMS:2012aa,Abelev:2014dsa},
        direct photon \cite{Chatrchyan:2012vq}, Z$^0$ \cite{Chatrchyan:2011ua},
        and W$^{\pm}$ \cite{Chatrchyan:2012nt} production at midrapidity. Figure from \cite{Abelev:2014dsa}.
        The next panel shows a zoom at low-\pt\ of \RpPb\ for identified particles ($\pi$, $K$,  p, $\Xi$). Figure from \cite{Grosse-Oetringhaus:2014sga}. 
        (Right)
        Elliptic flow coefficient $v_2$ of charged particles \cite{Abelev:2012di, Chatrchyan:2012xq} measured by ALICE (orange) and CMS (green), compared to full jets from ATLAS (blue) \cite{Aad:2013sla} comprising both charged and neutral fragments 
        and charged jets $v_2^{jet}$ from ALICE (black) \cite{Adam:2015mda}. Note that the same parton \pt\ corresponds to different single particle, full jet and charged jet \pt. 
         Figure from \cite{Adam:2015mda}.  
         }
    \label{fig:RAA} 
\end{figure}

A summary of measurements of the \RAA\ 
of charged hadrons and electro-weak bosons 
is shown in Fig.~\ref{fig:RAA}-left;
the charged-particle \RAA\
in central \PbPb\ collisions at \sqrsNN\ 2.76~TeV \cite{ALICE:2012mj,Abelev:2014dsa,CMS:2012aa}
is compared with the \RAA\ of \W, \Z\ (from leptonic decays) \cite{Chatrchyan:2012nt,Chatrchyan:2011ua,Chatrchyan:2012xq,ATLAS:2013cta}
and (isolated) photons \cite{Chatrchyan:2012vq} 
at the same energy, 
as well as to the \RpPb\ from \pPb\ collisions at \sqrsNN\ 5.02~TeV \cite{Abelev:2014dsa}.

The LHC experiments also measured the \pT\ dependence of \RAA\ for different collision centralities \cite{CMS:2012aa, Abelev:2012hxa, Aad:2015wga}. 
The agreement of the different experimental measurements (when results are compared at similar centralities and rapidity windows) is remarkable. 
Charged-particle production is, as expected, progressively less suppressed going from central to peripheral \PbPb\ collisions. This observation is consistent with the predicted path-length dependence of energy loss. 
By comparing different models to experimental data from CMS \cite{CMS:2012aa} and ALICE \cite{Abelev:2012hxa} values of the transport coefficient $\hat{q} \approx $ 1.7--1.9  GeV$^2/c$ were extracted \cite{Burke:2013yra, Liu:2015vna}. 

In addition, the measurements (Fig.~\ref{fig:RAA}-left)  show that (isolated) photons and the \W\ and \Z\ bosons (in the leptonic decay channel) which do not carry colour charge, are not suppressed  \cite{Chatrchyan:2012nt,Chatrchyan:2011ua,CMS:2013rea}. This observation is compatible with
the hypothesis that the origin of the suppression of charged hadrons is the final-state
interactions with the created hot and dense medium.
Additional support comes from the \pPb\ data, 
which were expected to distinguish initial- from final-state effects.
First results of \RpPb\ measurements  from the \pPb\ pilot run at \sqrsNN\ 5.02~TeV \cite{ALICE:2012mj,Abelev:2014dsa} are also compared to the \PbPb\ data in  Fig.~\ref{fig:RAA}-left.
The ALICE  \RpPb\ measurement at high \Pt\ is comparable with unity and thus shows no indication of nuclear matter modification of hadron production 
and is consistent with 
binary collision scaling, as well as with \PbPb\ observables that are not affected by hot QCD matter, like direct photons \cite{Chatrchyan:2012vq}
and electroweak gauge bosons \cite{Chatrchyan:2011ua, Chatrchyan:2012nt}.
This is in line with the hypothesis that 
the observed suppression of hadron production at high \Pt\ in central \PbPb\ collisions
is not due to initial-state effects
and implies that the origin of this suppression is  
the produced hot quark-gluon matter in \PbPb\ collisions
\cite{ALICE:2012mj,Abelev:2014dsa}.

At the LHC, the nuclear suppression factor is also studied with identified particles,
which further constrains theoretical models of energy loss.
For light-flavoured particles, $\pi$, $K$, p, at low \pt, a mass ordering related to radial flow is seen  \cite{Abelev:2014pua}, while at high \pt\ the \RAA\ of all particles is compatible with each other, showing that at high \pt\ the medium affects them in a similar way.
In addition, systematic studies of the suppression of heavy-flavoured particles
compared to light hadrons allow testing the predicted flavour-dependent hierarchy pattern of suppression, as detailed in Sec.~\ref{sec:heavy_intro} and Sec.~\ref{sec:HeavyFlavours}.
Moreover, studies of identified particles were also performed in \pPb\ collisions \cite{Grosse-Oetringhaus:2014sga}, see Fig~\ref{fig:RAA}-middle. The \RpPb\ for $\pi$, $K$, p  is compatible with unity for large \pT\ ($>$ 8 GeV/$c$) further supporting the view that the observed suppression is a final-state effect. At intermediate \pT\ a hint of mass ordering  (enhancement for p, $\Xi$) is visible. Such an enhancement, observed in \PbPb\ collisions, was associated to collective effects (see~\cite{P1}).  
This similar trend, also observed in \pPb\ collisions, suggests by analogy a possible collective origin.


\section{Jets}
\label{sec:ExpResultsHeavy_Jets}

\paragraph{a. Single jets}

The \RAA\ suppression pattern observed for charged particles was confirmed and extended using fully reconstructed jets over a wide \Pt\ range.
Compared to measurements based on individual high-\pt\ hadrons as described in Sec.~\ref{sec:ExpResultsHeavy_HighPt}, studies of reconstructed jets provide the advantage of a more direct connection between the \pT\ and direction of the measured jet and the ones of the initial parton.
 
In heavy-ion experiments at collider energies, jets can be reconstructed using a combination of tracking of charged particles with measurements in electromagnetic and hadronic calorimeters.
Typically the detected particles are grouped within a given angular region, i.e. a cone with radius $R$
which has to be optimized taking into account the background of the underlying event 
(formed by the soft particles produced in the collision).
Detailed studies have shown that 
it is possible to reconstruct jets above the fluctuations from the background event even in the high-multiplicity environment of heavy-ion collisions. 
Since the background can mimic medium-induced effects and affects the measurements, in particular correlation results \cite{Cacciari:2011tm}, increasingly sophisticated methods are being developed to control it \cite{Abelev:2012ej}.

The ATLAS jet measurements of \RAA\ \cite{Aad:2014bxa}
reveal that the strong observed charged-hadron suppression of \RAA\ of 0.5 in central \PbPb\ collisions at LHC persists up  to the highest measured \pT, extended up to 400 GeV/$c$, showing that the medium created in \PbPb\ collisions is so opaque that it can quench even the most energetic jets. 
It is, however, interesting to experimentally verify if very high-\pt\ jets should remain unaffected  and determine the related \pt\ range. 
In addition, a clear centrality dependence is observed, as for single hadrons. Also, \RAA\ shows a slow increase with \pT\ for central collisions. Such a rise could indicate a preferential quenching of gluon jets (relative to the quark jets), as the quark-to-gluon ratio increases with \pT. However, this ratio is also expected to increase with rapidity, and no such dependence is observed \cite{Aad:2014bxa}.
Further investigations \cite{Zapp:2011ek} show that the increase in single hadron \RAA\ is solely due to the initial reference spectrum from \pp\ collisions. 
Related 
measurements were performed by CMS \cite{CMS:2012rba} and  ALICE \cite{Adam:2015ewa}, the latter also accessing the particularly interesting lower \pT\ regions, down to \pt\ $\approx$ 30--40 GeV/$c$, which is the region where medium effects  and different processes are reflected and can be disentangled.
The results of the different LHC experiments are in good agreement despite the different analysis methods.
The observation that the single inclusive jet suppression is similar to that for single hadrons
could be understood if the dominant mechanism of parton energy loss is through radiation outside the jet cone used for the jet reconstruction.


To probe the energy loss suffered by the parton as a function of the distance traversed in the medium, 
and test the predicted path-length dependence of energy loss,
measurements of the variation of the jet yield in- and out-of-plane, 
employing \vtwo\ (see Sec.~\ref{sec:heavy_intro}), are performed.
Results on the jet \vtwo\ measurement \cite{Adam:2015mda,Aad:2013sla} are shown in Fig.~\ref{fig:RAA}-right.
A significant positive $v_{2}^{ch jet}$ is observed in semi-central collisions (black points and blue squares) while no (significant) \pT\ dependence is visible. 
This experimental observation establishes a clear relationship between the measured jet suppression and the details of the initial nuclear geometry; thus, it confirms not only the existence of the medium, but also the expectation that the jet suppression is strongest in the out-of-plane direction where partons traverse the largest amount of hot and dense matter.

\paragraph{b. Jet correlations} 

\begin{figure}[htb!]
\hspace{0.3cm}
\includegraphics[width=0.33\textwidth]{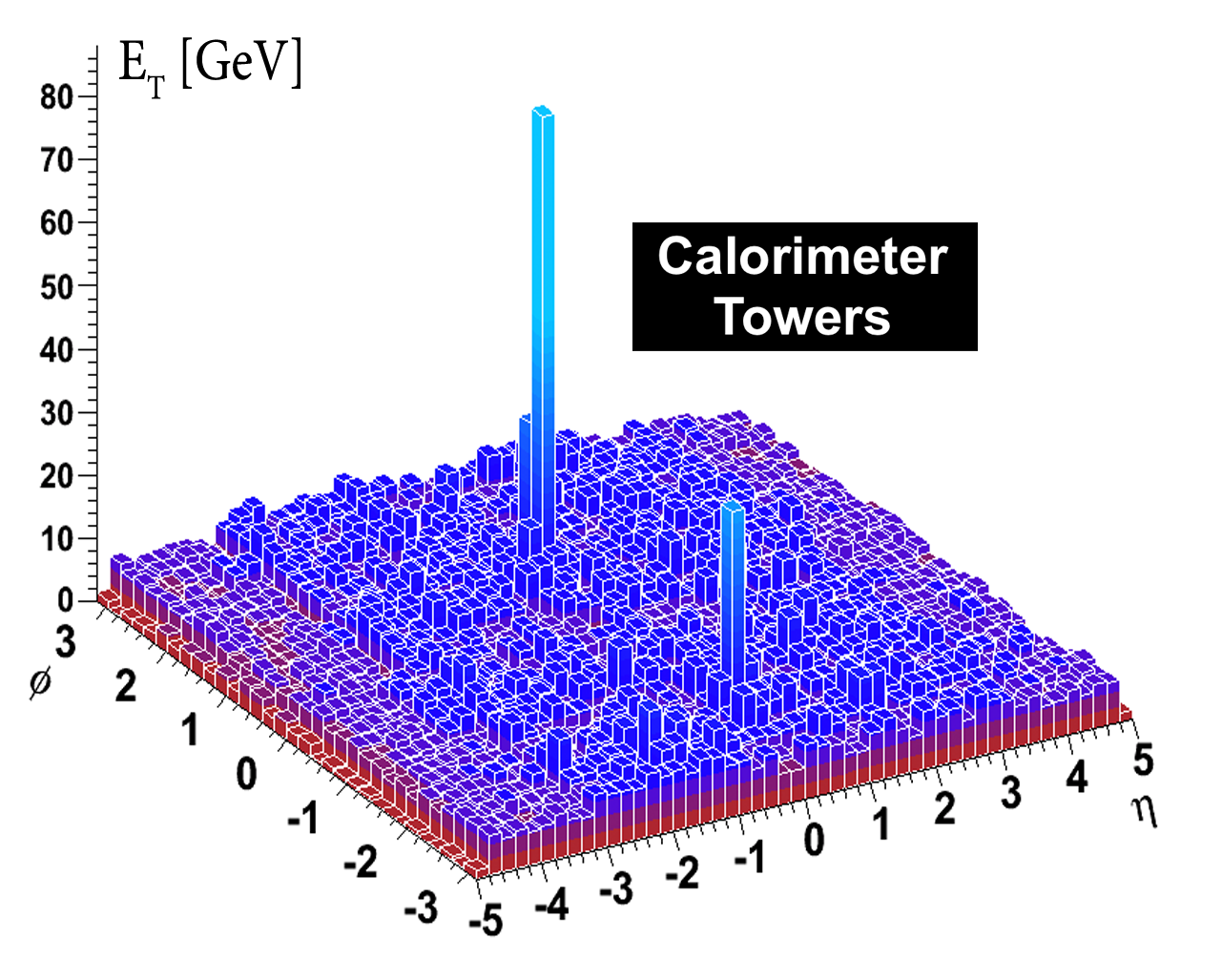}
\put(-165,105){(a)}
\includegraphics[width=0.66\textwidth]{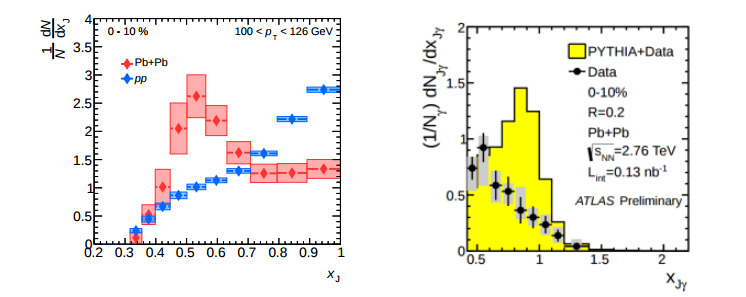}
\put(-310,105){(b)}
\put(-155,105){(c)}
\caption{ (a) Energy distribution showing not fully balanced (``asymmetric") jets and soft particles at large angles registered by ATLAS for \PbPb\ collisions at \sqrsNN\ 2.76 TeV. Figure from \cite{ATLASEventDisplays}.
(b) The dijet asymmetry distribution for the 0--10\% centrality bin (red) and pp (blue). Figure from \cite{ATLAS-CONF-2015-052}.
(c) The $x_{\rm{J\gamma}}$ distribution for \PbPb\ collision data (closed symbols) compared with PYTHIA simulation of the true jet/true photon distributions (yellow histogram). Figure from \cite{ATLAS:2012cna}.
}
\label{fig:Dijets_ATLAS}
\end{figure}

Even inspecting by bare eye the energy distributions of the first heavy-ion events recorded at LHC,  see Fig.~\ref{fig:Dijets_ATLAS}-(a), one could observe a large number of events with a high-\pT\ reconstructed jet (e.g. of order of 100 GeV/$c$) whose energy was not fully balanced by the energy of a back-to-back high-\pT\ partner jet.

This modification of the dijet properties relative to the reference pp collisions was quantified by measurements of the average dijet asymmetry for the leading and sub-leading jets in the event
\cite{ATLAS:2015lla, ATLAS-CONF-2015-052}, $x_{\rm J}=p_{\rm T}^{Jet 1}/p_{\rm T}^{Jet 2}$, 
which is shown in Fig.~\ref{fig:Dijets_ATLAS}-(b), 
for the most central \PbPb\ collisions.
A large number of unbalanced dijets is observed. Moreover, 
data show a strong evolution of the shape of the dijet asymmetry distribution as a function of both centrality and \pT .
These experimental observations can be understood assuming that the back-to-back partons traverse different path lengths in the QGP medium, and hence suffer different energy loss.
A sizeable difference between the \PbPb\ and \pp\ reference distributions persists even for the highest \pT\ range, demonstrating that the medium created in \PbPb\ collisions can indeed quench also jets with very high \pT . 
In addition, the azimuthal angle correlations of dijets was studied in \pp\ and \PbPb\ collisions for different centralities \cite{Chatrchyan:2011sx, ATLAS-CONF-2015-052}, but no strong modification of azimuthal correlations was observed, indicating that the energy loss suffered does not alter the azimuthal angle of the back-to-back partners relative to the pp (vacuum) distributions. 

Studies of \pPb\ collisions show that for minimum bias events the nuclear modification of dijet \cite{Chatrchyan:2014hqa} and single jet distributions \cite{ATLAS:2014cpa} 
is very small and compatible with expectations from the nuclear modification of parton densities 
(in fact these measurements can be used to constrain them \cite{Paukkunen:2014pha}).

Measurements of inclusive jets and dijets  provide only limited information since the initial jet energy is not well defined. In general, both initial partons of dijets suffer some energy loss, depending on the path length they traverse in the medium. Even if one of the jets is produced near the surface, the quantitative interpretation of the measurements of the dijet asymmetry is still biased from ambiguities (so-called ``surface bias"), related to the unknown absolute initial energy of the jet.
However, electro-weak gauge bosons retain the kinematics of the initial hard scattering because, not carrying colour charge, they are not affected by the medium \cite{Chatrchyan:2011ua,Chatrchyan:2012nt,Chatrchyan:2012vq}, as already shown in Fig.~\ref{fig:RAA}-left. 
Therefore, measurements of photon-jet events should not be affected by such ambiguities \cite{Wang:1996yh}. The photon can determine,  on an event-by-event basis, 
the initial direction and momentum of the back-to-back associated parton.
The measured energy of the reconstructed jet, compared to the energy of the photon,
should then better quantify the amount of the energy lost by the jet traversing the medium.
This makes the studies of photon-jet correlations one of the key methods
to determine the initial energy of the parent parton which generated the jet
\cite{Chatrchyan:2012gt, ATLAS:2012cna}.

Figure~\ref{fig:Dijets_ATLAS}-(c)  presents
the mean fractional energy distribution carried by the jet opposite an isolated photon, $x_{J\gamma}=p_{\rm T}^{Jet}/p_{\rm T}^{\gamma}$,
in \PbPb\ collisions
compared to simulations (yellow histogram) \cite{ATLAS:2012cna}.
With increasing centrality the distribution of collision data is seen to shift toward smaller $x_{\rm{J\gamma}}$,
which suggests that more and more of the jet momentum distribution is found below a minimum
$x_{\rm J\gamma}$,
in contrast to the MC data, where the distribution of the ratio of the ``true jet'' to the ``true photon''
shows no centrality dependence.
The photon-jet studies provide a clear evidence of parton energy loss,
reducing biases present in charged-hadrons measurement, because of a better determination of the photon initial energy. Similar ongoing measurements of $Z$-jets promise to provide an additional handle on such studies. Precision studies involving a variety of observables and higher statistics are expected to improve the accuracy of these measurements and to allow a precise determination of the absolute parton energy loss as a function of the parton \pT\ and the average path length traversed in the medium. 

\paragraph{c. Internal structure of jets} 

In addition to the studies quantifying the energy loss due to the parton interaction with the medium 
it is also interesting to study the 
modifications of the jet fragmentation properties with respect to the fragmentation in \pp\ collisions. 
These are usually studied using jet fragmentation functions defined as the yield of fragments in bins of fractional reconstructed jet momentum $z = p_{\rm T}^{\rm track} / p_{\rm T}^{\rm jet}$. 
In \PbPb\ collisions at \mbox{\pT\ $> 4$ GeV/$c$}
no significant modification of the fragmentation function is observed \cite{Chatrchyan:2012gw}.
However, if lower \pt\ particles are included in the analysis, the fragmentation function in \PbPb\ collisions \cite{Chatrchyan:2012gw,  Chatrchyan:2014ava, ATLAS:1395333} 
shows an enhancement of soft particles, a suppression of particles with intermediate momentum fractions, and little modification of hard ones. 
While the excess of soft particles is related to the jet quenching, the intermediate and high-\pt\ jet structures may also be explained by an increasing gluon-to-quark ratio. See \cite{Spousta:2015fca} for a detailed discussion.

Currently, the complexity of the measurements and the many biases that accompany them make a direct comparison between experimental and theoretical
observables very difficult. New observables are being established that are both measurable and calculable with well-controlled
precision. An example of such new analyses is the measurement
of the modification of the jet shapes, as proposed by CMS \cite{Chatrchyan:2013kwa}. This measurement indicate a redistribution of the energy inside the cone in central \PbPb\ collisions. 
Specifically, the results show a depletion of a fraction of the jets' \pt\ at intermediate radii, $0.1 < r < 0.2$  and an excess at large radii, $r > 0.2$.
Another measurement, of the radial energy profile of the jet 
\cite{Cunqueiro:2015dmx} proposed by ALICE, allows discriminating between two competing scenarios: jet quenching that results in intra-jet broadening (due to the energy loss the jet cone becomes wider), or collimation (most of the energy carried by the jet is collimated closer to the jet axis).
The measurements indicate that the jet cores in \PbPb\ are more collimated and have higher \pt\ than the jet cores that were simulated with models that did not include energy loss mechanisms.  

In addition to probing the internal structure of jets with reconstructed jets one can use two-particle correlation techniques. In particular, detailed studies of the baryon/meson yields in jets \cite{Veldhoen:2012ge} at low and intermediate \pt\ provide further insight into the jets composition. 
The measurements have shown no strong modification of the baryon/meson ratio relative to that in \pp\ collisions. This demonstrates that the overall baryon/meson enhancement observed in \PbPb\ collisions (see Ref.~\cite{P1}) is due to the bulk underlying event. 
Further ongoing measurements in ALICE based on strange particle reconstruction inside the jets promise to provide 
additional input to study, with higher precision, hadronchemistry in the jets and the question of the modification of the fragmentation function involving particle identification.

\paragraph{d. Energy flow outside of jets} 
A fraction of the ``lost" energy can be recovered within radii in the range $R=0.2$--$0.5$ from the jet axis  \cite{Chatrchyan:2013kwa}; however, even for the largest radius used for jet reconstruction at the LHC, a large suppression of inclusive jet yields and large dijet asymmetries are observed.
Therefore, a number of measurements is performed to track the flow of energy far outside the nominal jet radius \cite{Khachatryan:2016erx, CMS:2015kca, Chatrchyan:2011sx, Khachatryan:2015lha}. 
Overall, studies of the energy flow in dijets show that energy balance is achieved at low  momenta and very large radial distances relative to the jet axis.

\vspace{0.3cm}

Overall, the available data at the LHC and  theoretical progress allowed a quantitative extraction of the jet
quenching parameter $\hat{q}$ in the deconfined QGP matter,  which has also been calculated in lattice QCD \cite{ALICE:2012ab}. Values of $\hat{q}$ of several GeV$^2$/fm are extracted from the systematic study of \cite{CMS:2012vxa}.
However, from the theoretical point of view, the observations of the different properties of reconstructed jets 
are challenging 
the standard description of jet quenching in terms of medium-induced gluon radiation.
Describing all of the data in this section will be important for the overall understanding of the phenomenon of jet quenching, which is being intensively studied theoretically in a variety of approaches. 
A complete summary of published results of jet measurements from the LHC is listed in Table~\ref{Tab-jets}.

\section{Heavy flavours}
\label{sec:HeavyFlavours}
Heavy quarks are produced through initial hard-scattering processes at time scales $\sim1/2m_{c,b}$ (of order of 0.07 fm for charm and 0.02 fm for beauty), shorter than the QGP formation time  ($\tau_0 \sim $ 0.1--1 fm/$c$), and therefore witness the whole medium evolution. 
Heavy-flavoured particles are usually 
classified into open heavy flavour (particles with non-zero charm or beauty quantum numbers, e.g. $D$ mesons are the lightest particles containing a $c$ quark, $B$ mesons are the lightest particles containing a $b$ quark) and hidden (closed) heavy flavour, i.e. quarkonia, bound states of $Q\bar{Q}$ pairs, see Sec.~\ref{sec:ExpResultsHeavy_Quarkonium}.

Unlike light quarks and gluons, that can be produced or annihilate during the entire evolution of the fireball, the annihilation rate of heavy quarks is small \cite{BraunMunzinger:2007tn}. 
In general, the total charm and beauty yields are not affected,
contrary to their phase-space distributions, 
which opens up the possibility to better quantify 
their modification due to their interactions with the traversed medium (see Sec.~\ref{sec:heavy_intro}).
Furthermore, their interaction with the medium  
may redistribute their momenta to lower values; 
therefore, they may thermalize in the system and participate in the collective flow dynamics.
Experimentally, two observables are usually studied to probe the interaction of heavy quarks with the medium;
namely, the nuclear modification factor, \RAA\, and the azimuthal anisotropy, quantified via the elliptic flow coefficient \vtwo.   They can exploit the mass and path-length dependence of heavy-quark energy loss and their comparison to theoretical models can give access to the measurement of the medium transport coefficients.
Indeed, first measurements at RHIC (with electrons from heavy-flavour decays) established that heavy quarks lose energy as they traverse the hot and dense medium created in heavy-ion collisions and participate in the collective expansion of the fireball \cite{Adare:2006nq}.

At LHC, taking advantage of the much larger heavy-flavour production cross sections, 
heavy-flavoured particles were measured systematically in all systems, \pp, \pPb, and \PbPb,
in many different channels, expanding the kinematic reach and increasing the precision of the measurements.
Such studies at LHC  were also  extended to heavy-flavoured tagged jets,  reconstructed in heavy-ion collisions for the first time.
Most importantly, the simultaneous measurement at the LHC of hidden and open heavy flavoured particles (which gives access to the total production cross section), should make it possible to better interpret the medium-induced effects (such as the observed \Jpsi\ suppression at SPS, where measurements of open charm were not accessible).

Here we present some selected results, also highlighting measurements of the reference systems that 
are relevant to the study of heavy flavours. 
The published results to date are summarized in Tables \ref{tab:OpenHeavy_expSummary_LHC} and \ref{tab:Quarkonium_expSummary_LHC}. Other review papers can be found in \cite{Andronic:2015wma,Andronic:2014zha,Averbeck:2013oga}.

\subsection{Open heavy flavours}
\label{sec:ExpResultsHeavy_HeavyFlav}

\begin{figure}[hbt]
    \includegraphics[width=0.49\textwidth,height=7cm]{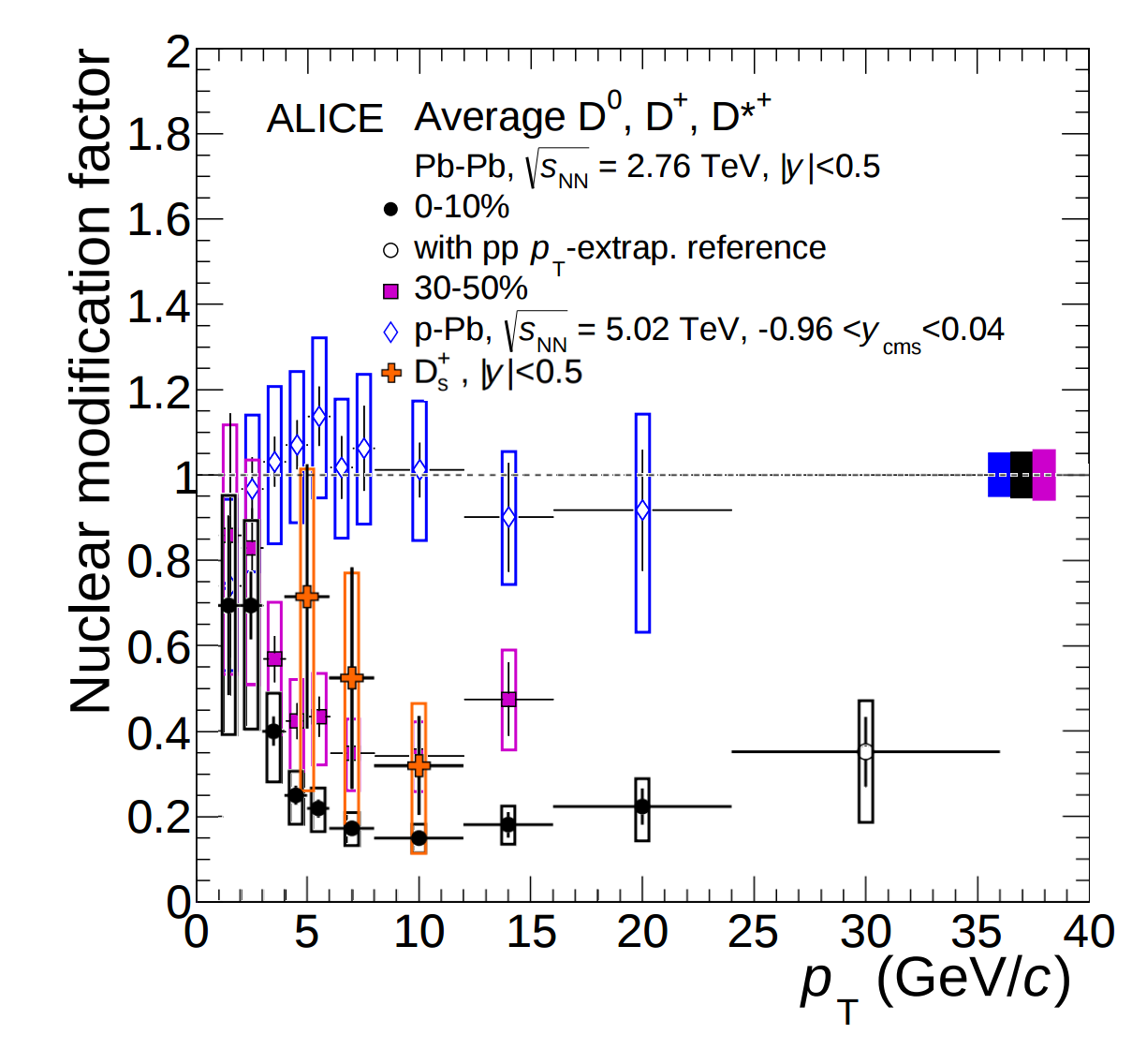}
    \hspace{0.5cm}
    \includegraphics[width=0.45\textwidth,height=7.3cm]{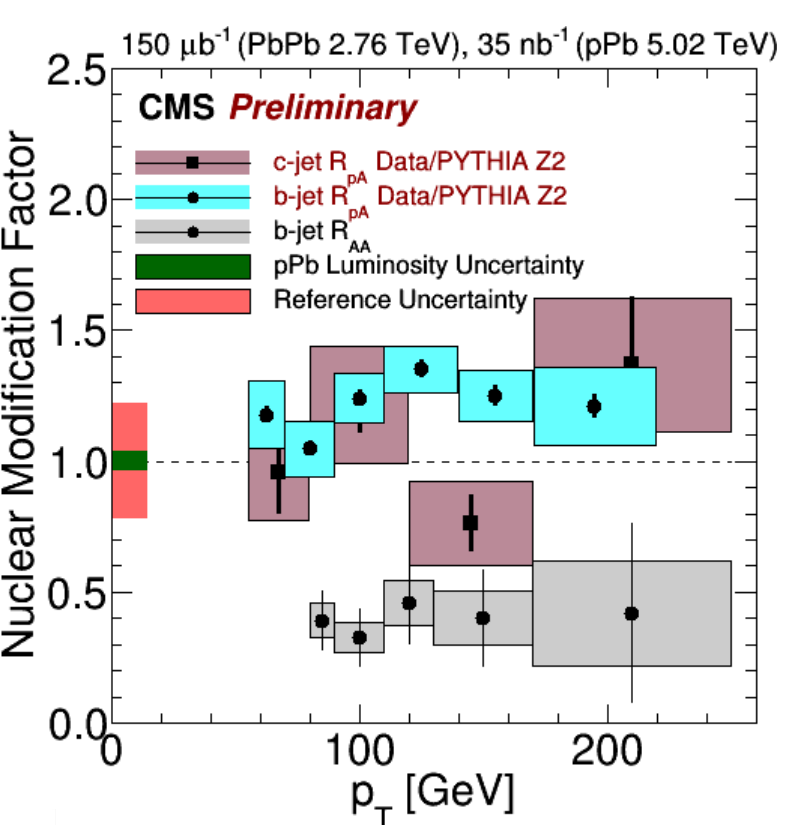} 
     \caption{
         (Left) Prompt $D$ meson \RAA\ (average of $D^0$ , $D^+$ and $D^{*+}$ \RAA ) \cite{Adam:2015sza} and
         prompt $D^{+}_{s}$ mesons \RAA\ \cite{Adam:2015jda}
         as a function of \pT\ in \PbPb\ collisions at \sqrsNN\ 2.76 TeV compared to 
         the prompt $D$ \RAA\ in \pPb\ collisions at \sqrsNN\ 5.02 TeV \cite{Abelev:2014hha}.
      (Right) 
      \RAA\ of $b$-jets in \PbPb\ at \sqrsNN\ 2.76 TeV \cite{Chatrchyan:2013exa},  $b$-jets \cite{Khachatryan:2015sva}, and $c$-jets in \pPb\  at \sqrsNN\ 5.02 TeV  \cite{CMS:2015lca} collisions. Figure from ~\cite{QM2015}.
         }
    \label{fig:RAA_D_and_jets}      
\end{figure}

\paragraph{a. D measurements}
Measurements of open charm mesons are used to determine the differential charm production cross section.
Figure~\ref{fig:RAA_D_and_jets}-left shows 
the \Pt\ dependence of the average \RAA\ of prompt $D$ mesons ($D^0$, $D^+$, and $D^{*+}$; the \RAA\ results are compatible \cite{delValle:2012qw}  and therefore they are often averaged)  in \PbPb\ collisions at \sqrsNN\ 2.76~TeV~\cite{ALICE:2012ab,Adam:2015nna,Adam:2015sza}
measured up to \Pt\ = 36 GeV/$c$, for two centrality intervals (0--10\% and 30--50\%).
At high-\Pt, the $D$ meson yield is strongly suppressed;  
for the most central collisions by a factor of about four at \Pt\ around 10 GeV/$c$. 
For more central events, an increase of the suppression is observed, compatible with the expected path-length dependence of energy loss.

The studies of the $D$ mesons family were complemented with the measurement of the $D_s$ meson which consists of a charm and an antistrange quark, and was measured in \PbPb\ collisions for the first time \cite{Innocenti:2012ds}. 
The first measurement of $D^+_s$, of very limited statistics, for the 10\% most central collisions, is also presented in Fig.~\ref{fig:RAA_D_and_jets}-left. 
The results, at high-\Pt\ \mbox{(8 $<$ \Pt\ $<$ 12 GeV/$c$)}, show, within the precision of the measurement,
a substantial suppression, compatible with that of non-strange mesons,
indicating  strong coupling of charm quarks 
with the deconfined created medium \cite{Adam:2015jda}. 
At lower \Pt\ (4 $<$ \Pt\ $<$ 8 GeV/$c$), 
within large uncertainties,
the $D^+_s$ \RAA\ is larger than the $D$-meson \RAA\ \cite{Adam:2015jda}. 
This is in agreement with the expectation that $D_s$ is sensitive to a possible hadronisation of charm quarks  via their (re)combination with light quarks from the medium.
Because of the predicted strangeness enhancement in the QGP, an increase of the $D_s$ \RAA\ relative to the other $D$-mesons is expected in the \Pt\ range where (re)combination may be relevant \cite{He:2012df}. 
Such measurements can contribute to a better understanding of the different underlying processes which is pursued studying in a systematic way the different collisions systems, \pp, \pPb, \PbPb.

The \RAA\ of $D$ mesons in minimum bias \pPb\ collisions at \sqrsNN\ 5.02 TeV is also compared to the \PbPb\ data in Fig.~\ref{fig:RAA_D_and_jets}-left.
It is found to be consistent with unity at high-\Pt\
which supports the hypothesis that the  suppression of the $D$ mesons yield observed in central \PbPb\ collisions is a final-state effect induced by the medium. 
%
%
Theoretical calculations including CNM effects, 
that could be present in the initial nuclei,
are in good agreement with the experimental results \cite{Abelev:2014hha}. 
The CNM effects are found to be small at intermediate and high \pt\ while they increase towards low \pt.

%
%
\vspace{-0.2cm}
\paragraph{b. $B$ measurements} 

\begin{figure}[t!]
    \includegraphics[width=0.5\textwidth]{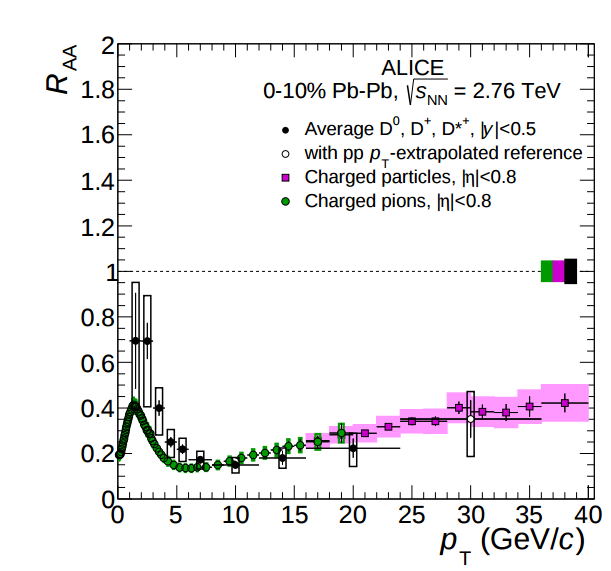}
    \hspace{0.5cm}
    \includegraphics[width=0.44\textwidth]{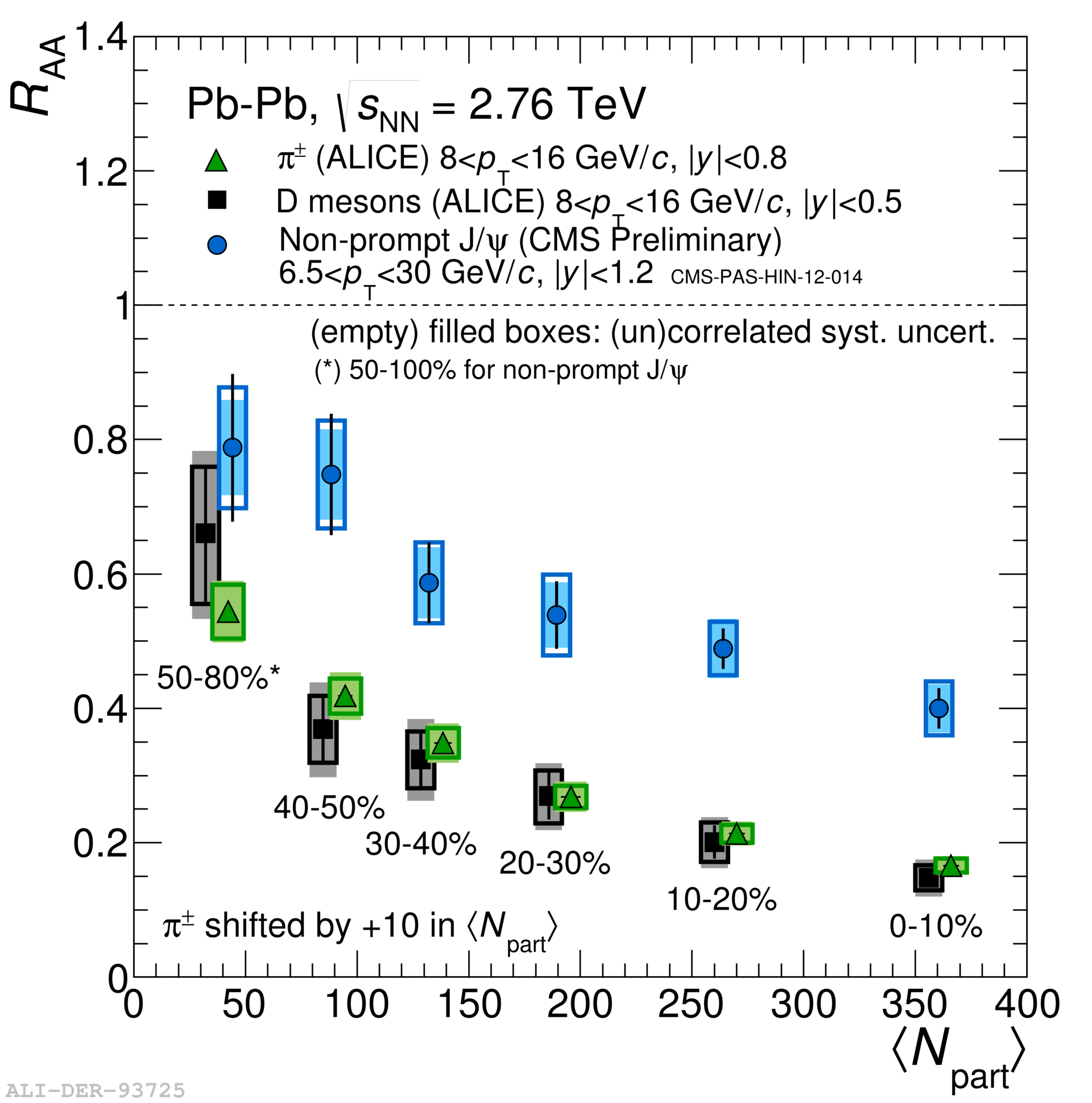}
    \caption{(Left)
    Prompt $D$-meson \RAA\ (average of $D^0$, $D^+$ and $D^{*+}$ \RAA) \cite{Adam:2015sza} compared to the $\pi$  \cite{Abelev:2014laa} and charged-particles \cite{Abelev:2012hxa} \RAA\  as a function of \pT\  for 0--10 \% centrality for \PbPb\ collisions at \sqrsNN\ 2.76 TeV. Figure from \cite{Adam:2015sza}.
    (Right) Charged particle (black squares), $D$ mesons (green triangles) \cite{Adam:2015nna} and
    non-prompt \Jpsi\ (blue circles) \cite{CMS-PAS-HIN-12-014}
    \RAA\ as function of centrality (\Npart ) for \PbPb\ collisions at \sqrsNN\ 2.76 TeV.}
    \label{fig:RAA_pT_Npart} 
\end{figure}

At the high LHC energies,
in addition to charm,
the beauty production can be measured with high-statistics. 
Measurements of beauty hadrons are typically exploiting 
decay channels that proceed as a $b$ to $c$ hadron cascade.
The first measurement of non-prompt \Jpsi\ (originating from $B$-mesons decays)
in heavy-ion collisions,  
exploiting the inclusive $B$ decays to  \Jpsi\ + X,
was performed by the CMS collaboration showing a significant suppression \cite{Chatrchyan:2012np}.
This first measurement was confirmed by  ALICE  \cite{Adam:2015rba}
and  was extended by CMS also exploring the \pT\ dependence of \RAA\  \cite{CMS-PAS-HIN-12-014}. 
Most recent results of non-prompt \Jpsi\ \RAA\
presented in Fig.~\ref{fig:RAA_pT_Npart}-right
as a function of participating nucleons, \Npart,
show a significant suppression of beauty 
for all measured centralities in the range 6.5 $<$ \pt\ $<$ 30 GeV/$c$.

Another method to study beauty is exploiting semi-leptonic decays of heavy flavour. 
Such an approach was employed by ALICE, first in pp \cite{Abelev:2014hla,Abelev:2012sca} and then in \PbPb\ \cite{Festanti:2014foa} collisions. The results of different methods are compatible, showing a similat suppression pattern for \Pt\ larger than about 5 GeV/$c$.
The measurements were extended with the study of 
correlations of electrons and the associated charged hadrons
exploiting specific characteristics of $B$ hadron decays \cite{Abelev:2014hla}. 

Complementary studies to those with $B$ mesons 
are based on measurements of reconstructed jets originating from $b$ quarks, 
which were extended, for the first time, from pp \cite{ATLAS:2011ac,Chatrchyan:2012dk} to heavy-ion collisions by CMS \cite{Chatrchyan:2013exa}, see Fig.~\ref{fig:RAA_D_and_jets}-right.
A sizeable suppression is observed in the range 80 $<$ \pt\ $<$ 250 GeV/$c$.
The corresponding measurements of $c$ and $b$ jets in \pPb\ collisions \cite{Khachatryan:2015sva} show a \RpA\ compatible with unity, see Fig.~\ref{fig:RAA_D_and_jets}-right, which indicates that the suppression measured in \PbPb\ collisions is not due to cold nuclear-matter effects.

%
%
\vspace{-0.2cm}
 \paragraph{c. Hierarchy of suppression}

To experimentally test the predicted hierarchy of suppression $R_{AA}^{\rm light}<R_{AA}^{D}<R_{AA}^{B}$ (see Sec.~\ref{sec:heavy_intro}),
the nuclear modification factors \RAA\ of light- and heavy-flavoured particles are compared
in Refs. \cite{Andronic:2014zha,Andronic:2015wma,ALICE:2012ab,Abelev:2012hxa,Adam:2015sza}, 
with gradually improved statistics and analysis methods, including identified pions \cite{Abelev:2014laa} and reconstructed $b$-tagged jets \cite{Chatrchyan:2013exa}. 
A caveat to keep in mind in such comparison is that the predicted mass hierarchy is expected to be more pronounced at \Pt\ comparable to the quark masses and should progressively fade away at higher \Pt, while at low \Pt\ collective phenomena may play a role. 
In addition, a number of effects that may alter the predicted suppression pattern
 have to be taken into account \cite{Djordjevic:2013pba}. These include the differences between the primordial spectral shapes of the produced partons and their fragmentation functions, which are harder for heavy quarks than for light quarks. 
Furthermore, at low-\Pt, light-flavoured particles are mainly produced via soft processes, in contrast to the heavy-flavoured hadrons.


As can be seen in Fig.~\ref{fig:RAA_pT_Npart}-left, a definite conclusion for the comparison of light- and charmed-flavoured particles \RAA\ needs further support from experimental data.
It is also clear that the \Pt\ range selected for comparisons 
strongly affects the results when they are presented as function of centrality
(reflecting contributions from different physics processes at the different \Pt\ regimes).
Figure~\ref{fig:RAA_pT_Npart}-right shows the \RAA\ of  charged pions \cite{Adam:2015nna}, $D$ mesons \cite{Adam:2015nna} 
and non-prompt \Jpsi\  \cite{CMS-PAS-HIN-12-014}
as function of centrality (expressed in terms of the average number of participating nucleons \Npart).
The results show that
the \RAA\ of $D$ mesons and charged pions
measured in the range 8 $<$ \Pt\ $<$ 16 GeV/$c$ 
are consistent; within uncertainties \RAA($D$)$\approx$\RAA($\pi$)
for all studied collision centralities  \cite{Adam:2015nna}.
This observed agreement is reproduced by models which include
the different fragmentation functions and shapes of the primordial \Pt\ distributions of the different parton types \cite{Djordjevic:2013pba},
in addition to the expected energy-loss hierarchy.

%
%

On the other hand, 
the comparison of the \RAA\ 
of $D$ mesons \cite{Adam:2015nna} 
and non-prompt \Jpsi\ originating from beauty hadron decays  \cite{CMS-PAS-HIN-12-014},
presented in Fig.~\ref{fig:RAA_pT_Npart}-right
as a function of centrality,
shows the expected suppression pattern \cite{Adam:2015nna}.
The observed pattern is reproduced by pQCD models including mass-dependent radiative and collisional energy loss \cite{Djordjevic:2014tka}.
However, 
further effects have to be taken into account, 
including detailed considerations of the kinematics 
and the most appropriate \Pt\ ranges for such comparisons 
(e.g. the \Pt\ of the \Jpsi\ is shifted relative to that of the parent $B$ meson, by about 2--3 GeV/$c$, in the \Pt\ range of the CMS measurement 
\cite{Andronic:2015wma}).

These studies are complemented with measurements of $b$-jet production at high \pt .  
The observed suppression is significant \cite{Chatrchyan:2013exa} and is qualitatively consistent with the one of inclusive jets \cite{CMS:2012kxa}  suggesting that, at high \pt , a large  flavour-dependent parton energy loss is unlikely. 
Although quark mass effects are not expected to play a role at this high \Pt\ region, 
the difference expected for the energy loss between quarks and gluons 
should become apparent as a difference in the \RAA\ for $b$- and inclusive jets as the latter are dominated by gluon jets up to very high \Pt.
Further considerations, taking into account details of possible production mechanisms (such as gluon splitting \cite{Khachatryan:2015sva}), are being pursued.

%
%
\vspace{-0.2cm}
\paragraph{d. Heavy-flavour elliptic flow} 
The large energy loss suffered by heavy quarks in the QGP is an indication of their “strong coupling” with the medium which is dominated by light quarks and gluons. If heavy quarks interact strongly with the medium, heavy-flavoured hadrons could inherit the medium azimuthal anisotropy, quantified by \vtwo, see Sec.~\ref{sec:heavy_intro}.

\begin{figure}[hbt!]
\includegraphics[width=0.49\textwidth]{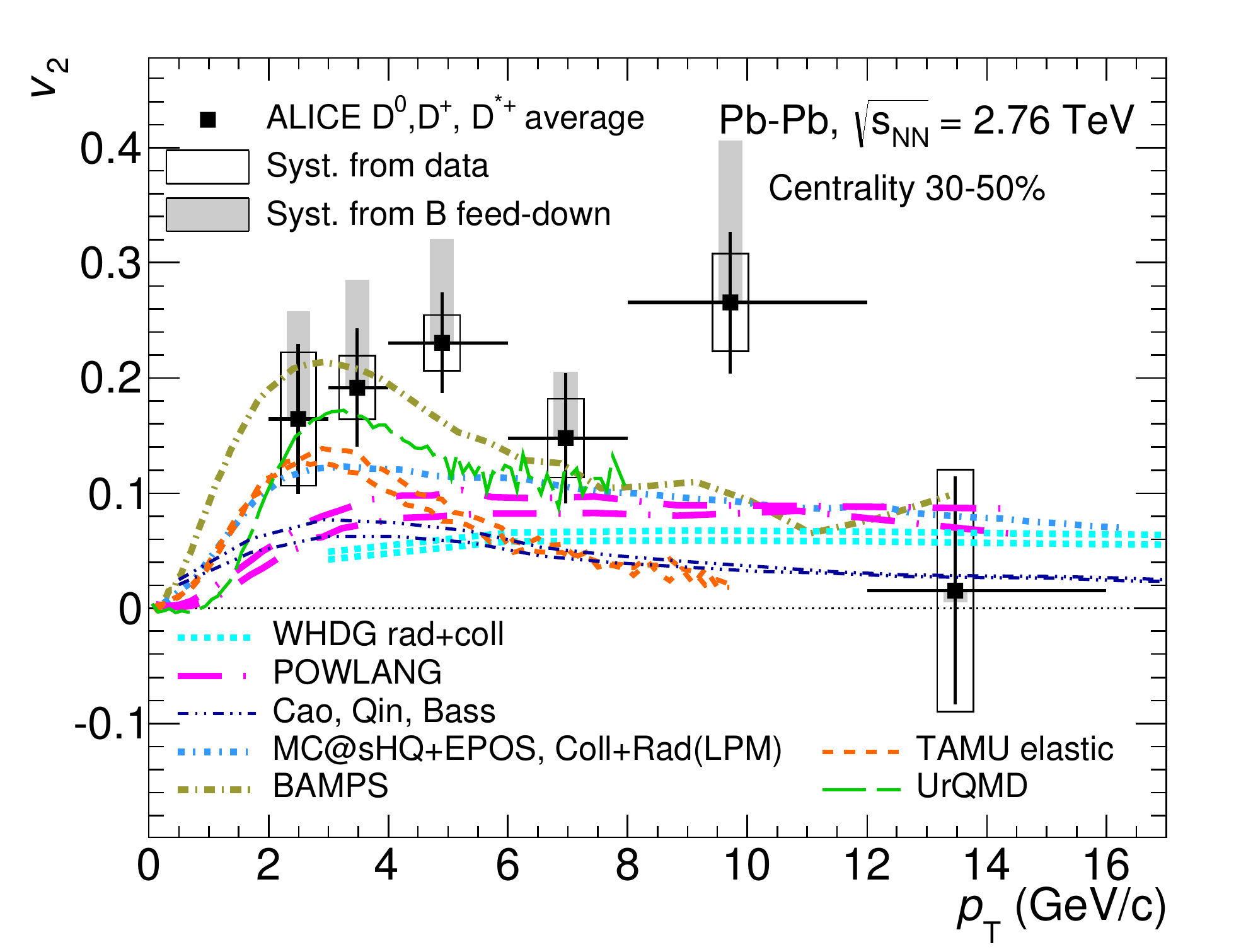}
\includegraphics[width=0.49\textwidth]{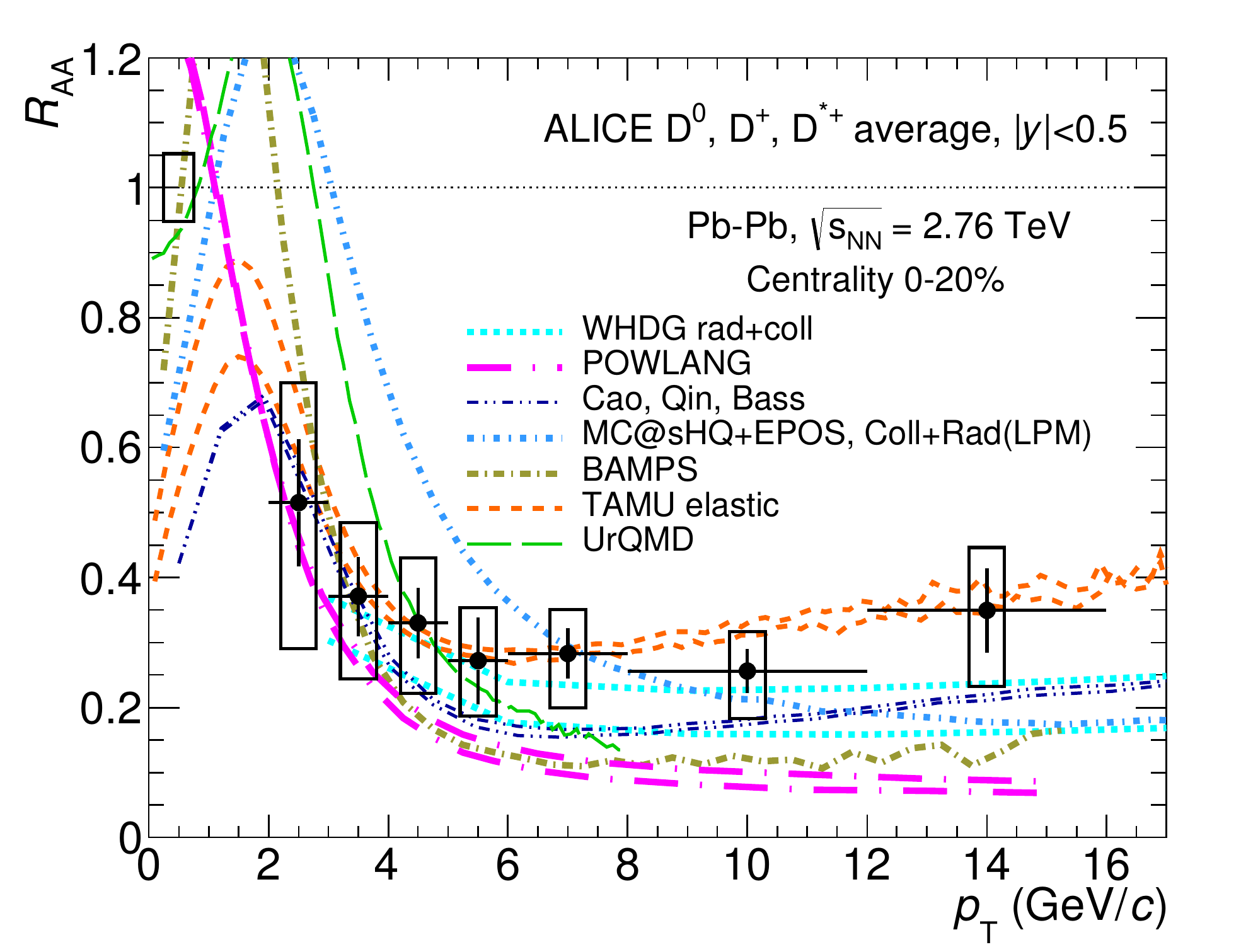}

    \caption{(Left) Averaged prompt $D$ mesons \vtwo\ for \PbPb\ collisions at \sqrsNN\ 2.76 TeV for the centrality range 30--50\% as a function of \pT\ \cite{Abelev:2014ipa}.
    (Right) Averaged prompt $D$ mesons  \RAA\  for \PbPb\ collisions at \sqrsNN\ 2.76 TeV for the centrality range 0--20\%  as a function of \pt\  \cite{Abelev:2014ipa}. 
     Both results are compared to theoretical models \cite{Horowitz:2011gd, Nahrgang:2013xaa, He:2014cla, Alberico:2011zy, Uphoff:2012gb, Lang:2012cx, Cao:2013ita}. 
    Figures from \cite{Armesto:2015ioy}.
}
    \label{fig:v2_D0s_PbPb}
\end{figure}

The ALICE collaboration studied the elliptic flow of charm in three centrality ranges. The averaged \vtwo\ values of  $D^0$, $D^+$, and $D^{*+}$ are presented in Fig.~\ref{fig:v2_D0s_PbPb}-left for the centrality range 30--50\%\ \cite{Abelev:2014ipa}.
These results represent the first direct observation of non-zero \vtwo\ of a heavy-flavoured particle. 
The $D$ meson results are compatible to the charged-particle \vtwo\ measurement obtained with the same analysis method, in the \pT\ range 2--8 GeV/$c$. 
Similar measurements of \Jpsi\ show a positive \vtwo\ (see Sec.~\ref{sec:ExpResultsHeavy_Quarkonium}) 
which can be used to disentangle the charmonium production mechanisms.

The large \vtwo\ of charm, at \pT\ around 2~GeV/$c$, of same magnitude as the light-hadron \vtwo\, can be considered as an indication of 
the charm-quark thermalization in the medium  which then also participate in the collective expansion. 
At higher \pT\, a positive \vtwo\ 
may be generated because of the difference of the path length in the medium for charm quarks emitted in-plane compared to those emitted out-of-plane, opening up the possibility to study the path-length dependence of the parton energy loss.
These observations confirm the significant interaction of heavy quarks with the medium.

Overall, the simultaneous measurement of \RAA\ and \vtwo\ provides a powerful tool to disentangle the interplay of various energy loss mechanisms and imposes important constraints on theoretical models. 
In Fig.~\ref{fig:v2_D0s_PbPb} the measured \vtwo\ and \RAA\ of  $D$ mesons are compared with different models.  Theoretical advances and systematic comparisons with experimental data focus on the  challenging task of describing quantitatively at the same time the \RAA\ and \vtwo\ of light- and heavy-flavoured particles over the full \pt\ range. 
In addition, such comparisons can give access to the heavy-quark transport coefficients in the QGP \cite{Adare:2006nq,Andronic:2014zha}.

\subsection{Quarkonia}
\label{sec:ExpResultsHeavy_Quarkonium}

\begin{figure}
\begin{center}
     \begin{minipage}{0.45\textwidth}
     \includegraphics[width=\textwidth]{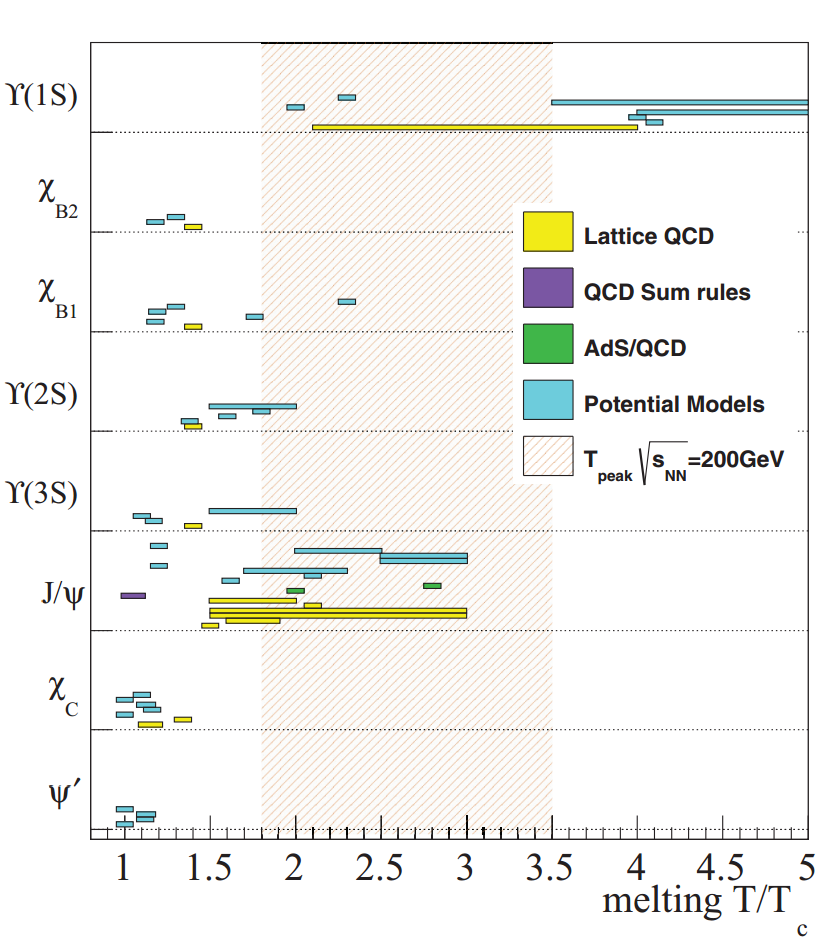} 
     \end{minipage}     
     \hspace{0.5cm}
     \begin{minipage}{0.45\textwidth}
\includegraphics[width=0.95\textwidth,height=3.8cm]{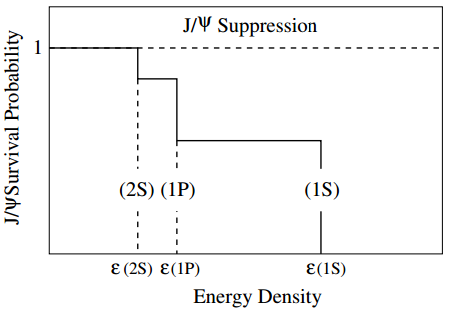}
\includegraphics[width=0.95\textwidth,height=4cm]{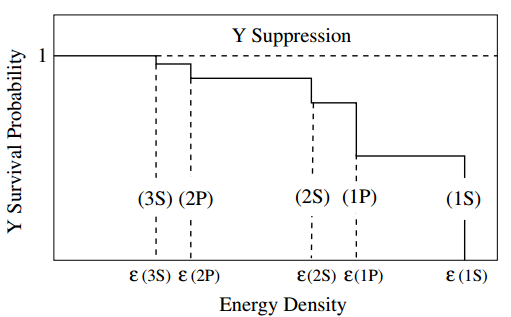}
     \end{minipage}

    \caption{
    (Left) Compilation of medium temperatures relative
            to the critical temperature ($T_c$), where quarkonium states are
            dissociated in the QGP. 
            Techniques used in calculations:
            lattice QCD \cite{Umeda:2000ym, Asakawa:2003re, Datta:2003ww, Jakovac:2006sf, Aarts:2007pk, Rothkopf:2011db, Aarts:2010ek, Aarts:2012ka, Aarts:2013kaa, Karsch:2012na, Adare:2014hje}, QCD sum rules \cite{Suzuki:2012ze, Morita:2007pt, Morita:2007hv, Song:2008bd, Morita:2009qk, Gubler:2011ua}, AdS/CFT \cite{Kim:2007rt, Fujita:2009wc, Noronha:2009da, Grigoryan:2010pj},
            effective field theories \cite{Brambilla:2008cx, Digal:2005ht}, and potential models \cite{Karsch:2012na, Alberico:2005xw, Mocsy:2007yj, Mocsy:2007jz, Petreczky:2010tk, Cabrera:2006wh, Riek:2010fk, Riek:2010py}. 
            Figure from \cite{Adare:2014hje}.
    (Right)  Sequential quarkonium suppression for \jpsi\ (upper) and \upsa (lower) states~\cite{Karsch:2005nk}.
    }
        \label{fig:quarkonium_intro_plots}
            \end{center}

\end{figure}

The nature and properties of the medium can be further studied 
exploiting measurements of quarkonia.
Quarkonium is a bound state of $Q\bar{Q}$, where $Q$ can be either a charm quark (forming a charmonium state) or a bottom quark (bottomonium state).
Such probes have played a special role since
it was argued that the disappearance of specific quarkonia states 
would signal the presence of a deconfined medium of a specific temperature.
In particular, the mechanism of \Jpsi\ ($c\bar{c}$) suppression in a deconfined medium, based on colour screening arguments (analogue to Debye screening in electromagnetic plasma) was first proposed in \cite{Matsui:1986dk} 
while further refinements predicted a pattern of ``sequential melting" 
\cite{Karsch:1990wi, Digal:2001ue, Karsch:2005nk}
dependent on the binding energy of quarkonium states, including both the $c\bar{c}$ and $b\bar{b}$ states. 
Because the Debye length in a deconfined system is temperature dependent \cite{Digal:2001ue}
the predicted hierarchy of quarkonium dissociation was thus expected to also probe the temperature of the medium, providing a so-called ``QGP thermometer" \cite{Digal:2001ue}.
In particular, calculations on the lattice give details on the screening mechanism and allow the calculation of the (static) colour screening length; 
for a review see \cite{Mocsy:2013syh}.
Table~\ref{tab:quarkonium_states}, derived from \cite{Satz:2005hx}, summarises the different $c\bar{c}$ and $b\bar{b}$  states and their binding energies  $\Delta E$ in the vacuum. The listed binding energies are the differences between the quarkonium masses and the open charm or beauty threshold, respectively. Figure~\ref{fig:quarkonium_intro_plots}-left 
summarises theoretical calculations of the ranges of the dissociation temperatures, relative to the critical temperature $T_{\rm c}$, of different bound states ($\psi'$ up to $\Upsilon$(1S)) obtained on the basis of different models. For example, on the basis of lattice QCD calculations, experimental observations of dissociation of the most bound states $\Upsilon$(1S)) would indicate a deconfined matter of a temperature in the range 2--4 $T_{\rm c}$ (350 -- 650 GeV).

\begin{table}[!b] 
  \caption{Charmonium and bottomonium states and their mass, binding energy  $\Delta E$ and radius. Table from \cite{Andronic:2015wma}.}
  \label{tab:quarkonium_states}

  \centering 
  \begin{tabular}{ccccccccc} 
    \hline 
    {\rm state}& $J/\psi$ & $\chic\text{(1P)}$ & $\psiP$ & 
    $\upsa$ & $\chib\text{(1P)}$ &  \upsb & $\chib\text{(2P)}$ & $\upsc$ \\ 
    \hline 
    {\rm mass~[GeV$/c^2$]}& 
    3.07 & 3.53 & 3.68 & 9.46 & 9.99 & 10.02 & 10.26 & 10.36 \\ 
    ${\rm binding}$ {\rm[GeV]}& 
    0.64 & 0.20 & 0.05 & 1.10 & 0.67 & 0.54 & 0.31 & 0.20 \\ 
    {\rm radius~[fm]}& 
    0.25 & 0.36 & 0.45 & 0.14 & 0.22 & 0.28 & 0.34 & 0.39 \\ 
    \hline 
  \end{tabular}
    
\end{table} 

The interpretation of the observed suppression pattern is not trivial; 
a quantitative description must consider feed-down from excited states, which contributes a significant fraction of the \Jpsi\ inclusive  yield in pp collisions.  Figure~\ref{fig:quarkonium_intro_plots}-right shows the sequential quarkonium suppression for \Jpsi\ (upper) and \upsa (lower) states. Furthermore, in addition to mechanisms related to hot matter, other effects related to cold nuclear matter, may affect the quarkonium production.
The assessment of the size of these effects is fundamental to interpret the AA quarkonium results.
Such CNM effects could include: (i) initial-state nuclear effects on the parton densities (shadowing); (ii) coherent energy loss consisting of initial-state parton energy loss and final-state energy loss; and (iii) final-state absorption by nucleons (expected to be negligible at LHC energies) \cite{Eskola:2009uj, Ferreiro:2013pua, Arleo:2012rs, Albacete:2013ei, Adeluyi:2013tuu, Chirilli:2012jd, Chirilli:2012sk, Arleo:2013zua}. 
The study of pA collisions is important to disentangle the effects of QGP 
from those of CNM, and to provide essential input to the understanding of nucleus-nucleus collisions.

In addition, at very high energies, a new production mechanism is thought to be at work (in the case of charmonium); 
namely, the abundant production of $c$ and $\bar{c}$ quarks\footnote{The number of $c\bar{c}$ pairs per event is increasing from 0.2 at SPS \sqrsNN\ 17.3~GeV to 10 at RHIC \sqrsNN\  200~GeV, and up to 85 at LHC \sqrsNN\ 2.76~TeV.} could lead to charmonium production by (re)combination of these charm quarks during the collision history  \cite{Thews:2000rj} or at hadronization \cite{BraunMunzinger:2000px,Andronic:2011yq}.

The measurements of \Jpsi\ in \PbPb\ collisions at  LHC was expected to provide an opportunity to disentangle dissociation and (re)combination effects. The observation of either one (or both) of these predicted phenomena i.e. quarkonium suppression or/and heavy-quark (re)combination implies the existence of a deconfined QGP state.

To interpret the results of quarkonia production and deduce the effects of a deconfined medium
it is important to understand if and how the medium presence modifies the fraction of produced $c\bar{c}$ pairs
that are going into charmonium formation.
The general idea at LHC is to normalize quarkonium production to the production of open charm that is dominant. 
While it is difficult to precisely quantify, the current understanding is that,  at first order,
the production process in elementary hadronic collisions starts with the formation of a $c\bar{c}$
pair which can then either lead to production of open charm (about 90\%) or bind to form a
charmonium state (about 10\% ) of all charmonia \cite{Satz:2013ama}.
Then, the crucial quantity to measure is the fraction of charmonia relative to open charm (and in general the fraction of quarkonia to the relevant open heavy-flavour  production \cite{Satz:1993pb, Satz:2013ama}).
If this quantity is measured over the full phase space, down to zero \Pt ,
then the kinematic biases and effects of any possible initial-state modification should cancel out.
Hence, any observed modifications relative to the pp collisions would then be due to final-state effects. 
Despite the fact that such measurements are experimentally challenging, first results from ongoing analyses at LHC are presented in \cite{Andronic:2015wma, Andronic:2014zha}; although the question of the appropriate \Pt\ intervals and interpretation of such comparisons are still to be addressed by theory.

\vspace{-0.2cm}
\paragraph{a. Charmonium results}

\begin{figure}[hbt]
\begin{minipage}{\textwidth}
\centering
\includegraphics[width=0.45\textwidth,height=5cm]{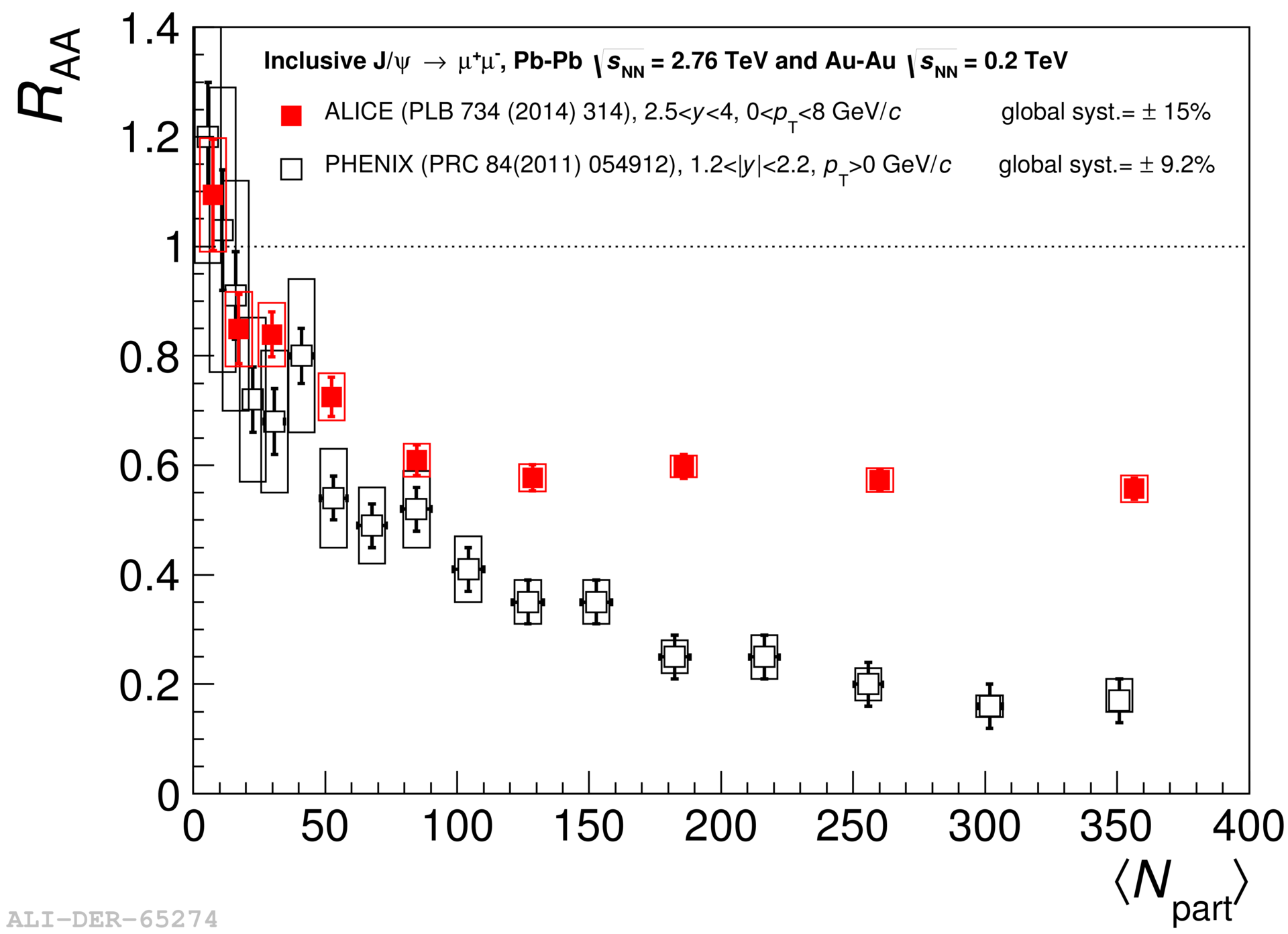}
\put(-220,130){(a)}
\hspace{0.4cm}
\includegraphics[width=0.46\textwidth,height=5cm]{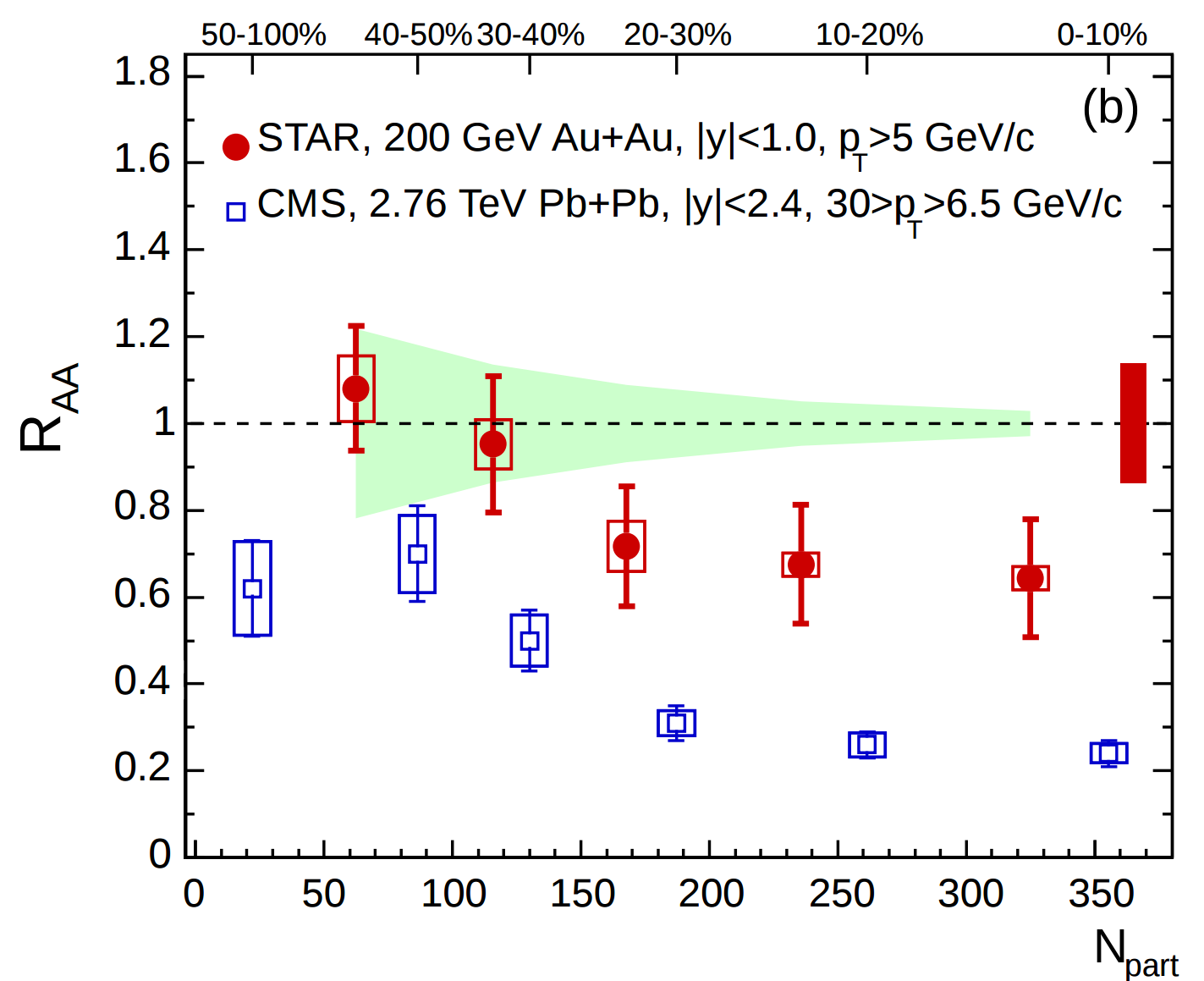}
\put(-220,130){(b)}
\end{minipage}
\begin{minipage}{\textwidth}
\centering
\includegraphics[width=0.45\textwidth]{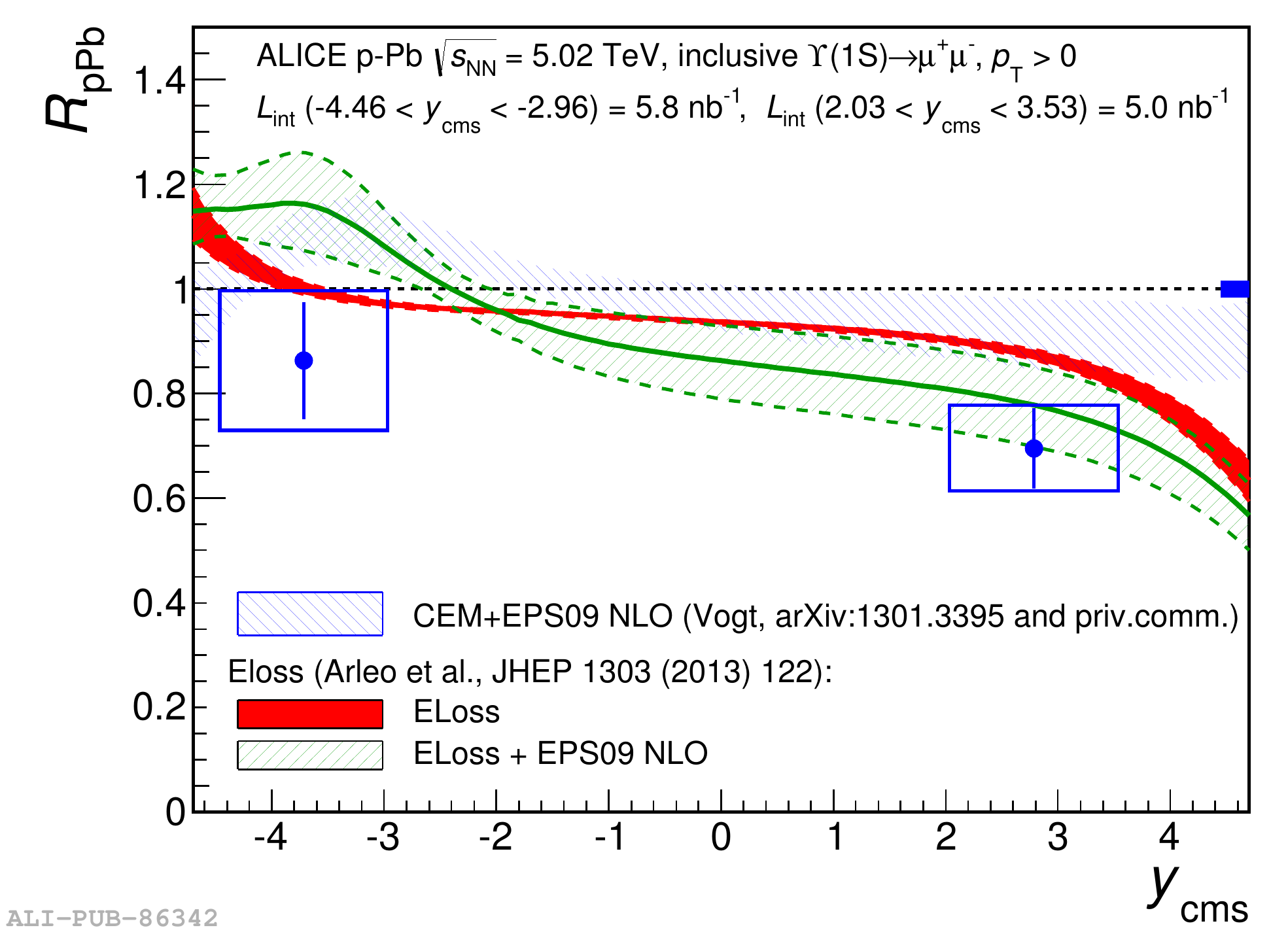}  
\put(-220,138){(c)}
\hspace{0.4cm}
\includegraphics[width=0.44\textwidth]{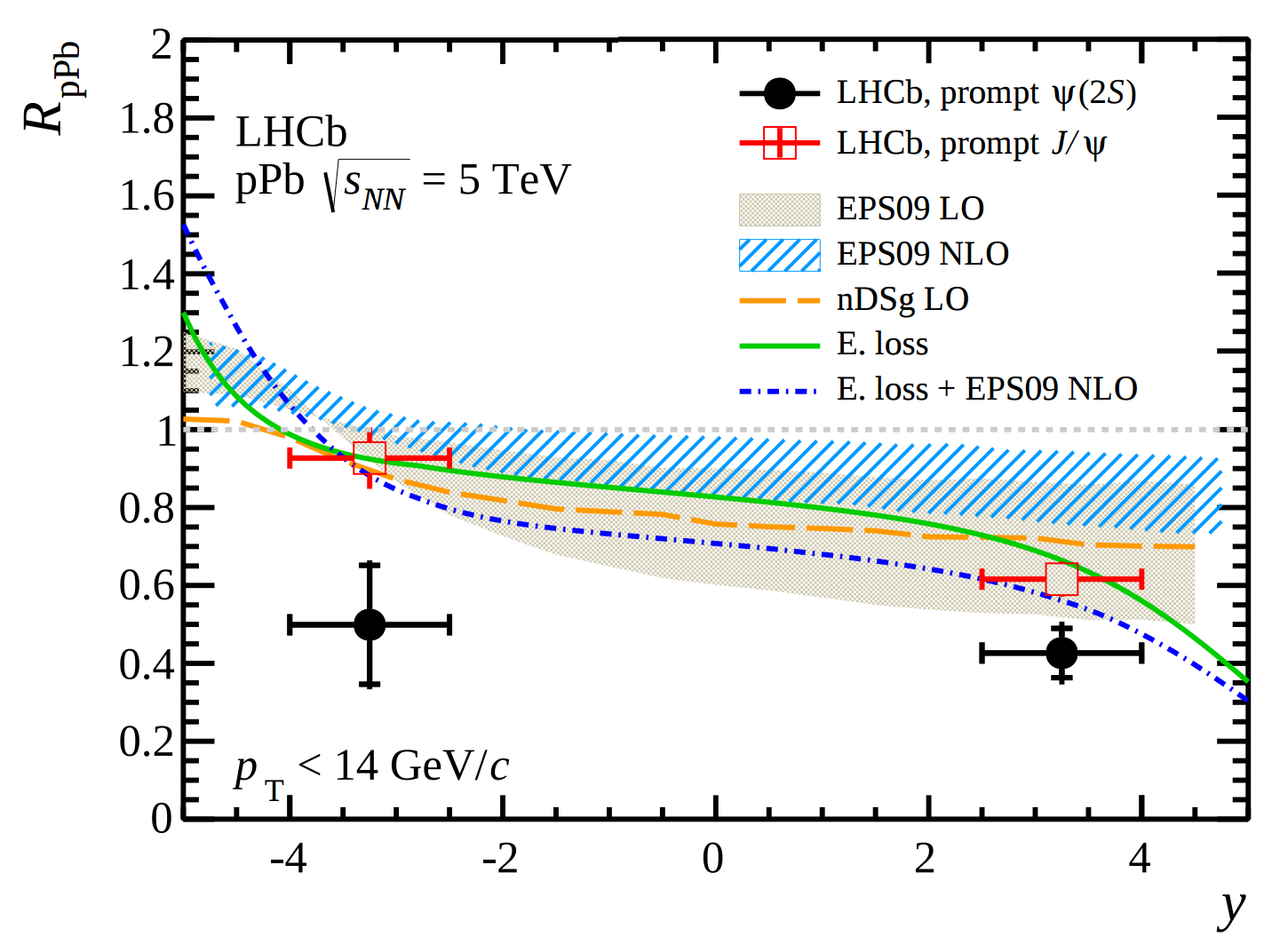}  
\put(-218,138){(d)}
\end{minipage}
\caption{(a) Centrality dependence of the nuclear modification factor \RAA\, of inclusive \Jpsi\ production in \PbPb\ collisions at \sqrsNN\ 2.76 TeV, measured at forward rapidity by ALICE compared to PHENIX-RHIC result in \AuAu\ collisions at \sqrsNN\ 200 GeV at low \pt . Figure from~\cite{Abelev:2013ila}. 
(b) Centrality dependence of the nuclear modification factor \RAA\, of \Jpsi\ production in \PbPb\ collisions at \sqrsNN\  2.76 TeV, measured at forward-rapidity by CMS compared to STAR-RHIC result in \AuAu\ collisions at \sqrsNN\  200 GeV at high \pt . Figure from~\cite{Adamczyk:2012ey}.
 (c) Nuclear modification factor of inclusive $\Upsilon$(1S) in \pPb\ collisions at \sqrsNN\ 5.02 TeV as a function of rapidity. Figure from \cite{Abelev:2014oea}.    
 (d) Nuclear modification factor \RpPb\ as a function of rapidity for prompt $\psi(2S)$ and \Jpsi\ 
    compared to the theoretical predictions from (yellow dashed line and brown
    band) \cite{Ferreiro:2013pua, delValle:2014wha}, (blue band) \cite{Albacete:2013ei}, and (green solid and blue dash-dotted lines) \cite{Arleo:2012rs}.
    Figure from \cite{Aaij:2016eyl}.
}
    \label{fig:JPsi_RAA}
\end{figure}

ALICE studied the evolution of the \Jpsi\ \RAA\ with centrality  for \pt\ $>0$ GeV/$c$ in \PbPb\ collisions at \sqrsNN\ 2.76~TeV, see Fig.~\ref{fig:JPsi_RAA}-(a), and  at \sqrsNN\ 5.02~TeV \cite{Adam:2016rdg}. These results are compared with RHIC measurements in \AuAu\ collisions at \sqrsNN\ 200 GeV and found to be strongly dependent on collision energy. 
Since more charm quarks are expected to be produced at the LHC than at RHIC, higher \RAA\ values at LHC can be attributed to the (re)combination which dominates \Jpsi\ production at low \pt. 
CMS results for high-\pt\ \Jpsi\ in the range $ 6.5 <$ \pt\ $ < 30$ GeV/$c$ compared with RHIC results at \pt\ $> 5$ GeV/$c$ show a behaviour opposite to that observed at low \pT, see Fig.~\ref{fig:JPsi_RAA}-(b). For those high-\pt\ particles the observed suppression is stronger at higher collision energy, as expected from the dissociation of the \Jpsi\ state due to the high temperature of the QGP.
Comparison with models has to take into account the contributions of the competing mechanisms of dissociation and (re)combination at low \Pt.

An additional experimental handle that is studied in order to disentangle the
different production mechanisms is the measurement of \vtwo.
A positive \vtwo\ (at low \pt ) measured for $D$ mesons (see Sec.~\ref{sec:ExpResultsHeavy_HeavyFlav}-d) 
indicates that charm quarks participate in the collective expansion of the QGP medium.
If \Jpsi\ is produced via (re)combination, it could inherit the elliptic flow of charm quarks in the QGP,
and  consequently, \Jpsi\ are expected to exhibit a measurable \vtwo.
ALICE measured a non-zero \vtwo\ for inclusive \Jpsi\ in semi-central \PbPb\ collisions at forward rapidity \cite{ALICE:2013xna} in the range 2 $<$ \Pt\ $<$6 GeV/$c$. Including statistical and systematic uncertainties the combined significance of a non-zero \vtwo\ in this \pT\ range is 2.7$\sigma$.
Transport models \cite{Zhao:2012gc, Liu:2009gx} that include a fraction of \Jpsi\ production through regeneration mechanisms
(at the level of about 30\%) describe fairly well the \vtwo\ measurement
(and at the same time describe the \Jpsi\ \RAA\ shown in Fig.~\ref{fig:JPsi_RAA}-(a--b)).
Furthermore, primordial \Jpsi\ may acquire a \vtwo\ component  induced by the path-length dependence of energy loss.
Thus, the final \vtwo\ with a predicted maximum at \Pt\ about 2.5 GeV/$c$, compatible with the ALICE measurement, could be the result of the interplay of the regeneration component, which dominates at low \Pt\ and the primordial \Jpsi\ component
which takes over at higher \Pt.
Complementary measurements of prompt \Jpsi\ \vtwo\ by CMS \cite{CMS:2013dla} cover also higher \Pt\ up to about 30 GeV/$c$
supporting the path-length dependence of partonic energy loss \cite{Chatrchyan:2012xq}.

\vspace{-0.2cm}
\paragraph{b. Charmonia in \pA\ collisions}

The production of the \Jpsi\ meson was studied using \pPb\ data by ALICE \cite{Abelev:2013yxa,Adam:2015iga,Adam:2015jsa} and LHCb \cite{Aaij:2013}.
The \Jpsi\ $R_{\rm pA}$ shows a strong dependence on rapidity and \pt , and the results are in agreement,  within uncertainties, with theoretical predictions, based on a pure nuclear-shadowing scenario \cite{Vogt:2010aa,Albacete:2013ei}, as well as with partonic energy loss, either in addition to shadowing or as the only nuclear effect \cite{Arleo:2013zua}. 
Similar measurement of inclusive $\Upsilon$(1S) \RpPb\ integrated over the backward or forward rapidity ranges, are compared to several model calculations
in Fig.~\ref{fig:JPsi_RAA}-(c). None of the calculations fully describe the data, both at the backward and forward rapidity. Additional measurements with higher statistics are needed to further constrain the models.

In addition, the study of the $\psi(2S)$ state \cite{Abelev:2014zpa,Aaij:2016eyl}, 
which is more weakly bound than the \Jpsi, can provide  further insight on charmonium behaviour in \pA\ collisions. 
At LHC energies, the time that the $c\bar{c}$ pair spends in the created medium is much shorter than the time the pair needs to evolve into a fully-formed resonance state, like the \Jpsi\ or the $\psi(2S)$. Thus, cold nuclear-matter effects, such as shadowing or coherent energy loss, should affect only the pre-resonant state and therefore are expected to be very similar for the two charmonium states.
The results 
for prompt $\psi(2S)$ and \Jpsi\ are shown in  Fig.~\ref{fig:JPsi_RAA}-(d). The $\psi(2S)$ suppression 
was unexpectedly found to be stronger than that of the \Jpsi\ \cite{Leoncino:2015ieu}.
Thus, an additional final-state mechanism, which affects only the weakly bound $\psi(2S)$, is needed to explain the observed pattern. 
A natural explanation is the introduction of the $\psi(2S)$  dissociation by comovers in a hadronic medium (possibly including a short-lived QGP phase created in \pA\ collisions)
\cite{Ferreiro:2014bia,Du:2015wha}.

\vspace{-0.2cm}
\paragraph{c. Bottomonia results}

Furthermore, the high LHC energies open up the possibility for high-statistics precision studies of the $\Upsilon$ family ($b\bar{b}$ bound states distinguished by the hierarchy of their binding energies). 
A distinct suppression pattern of the $\Upsilon$ (nS) states is expected, within an in-medium dissociation scenario, which would reflect their different binding energies, e.g $\Upsilon(1S)$ should melt at higher energy than $\Upsilon(2S)$ and $\Upsilon(3S)$ as was shown in Table \ref{tab:quarkonium_states} and Fig.~\ref{fig:quarkonium_intro_plots}-right.
An added advantage is that, for the bottomonium family, the uncertainties due to CNM effects as well as (re)combination effects are expected to be of less importance compared to the charmonium family \cite{Andronic:2015wma}.

\begin{figure}[hbt]
\includegraphics[width=0.33\textwidth]{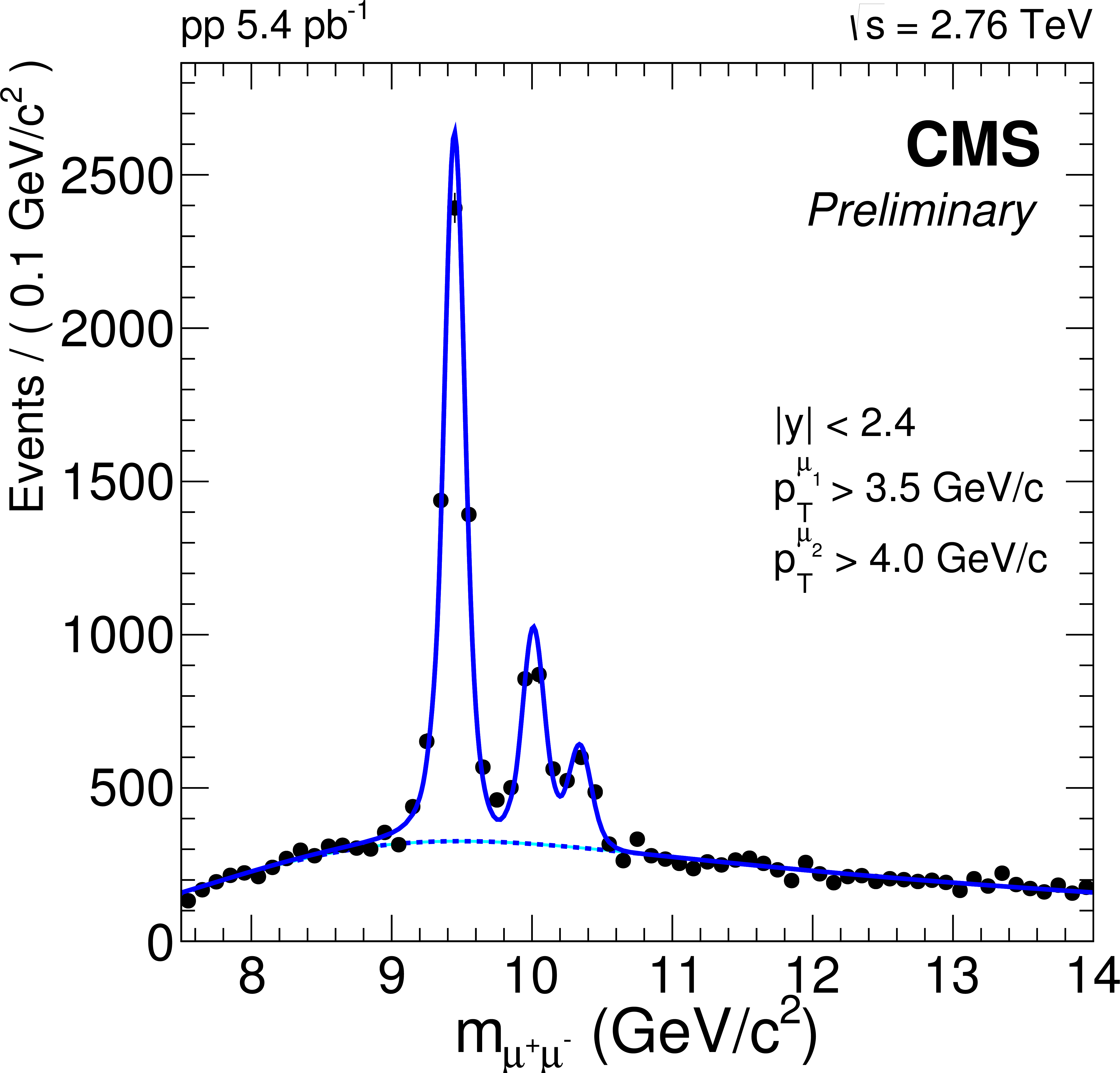}    \includegraphics[width=0.33\textwidth]{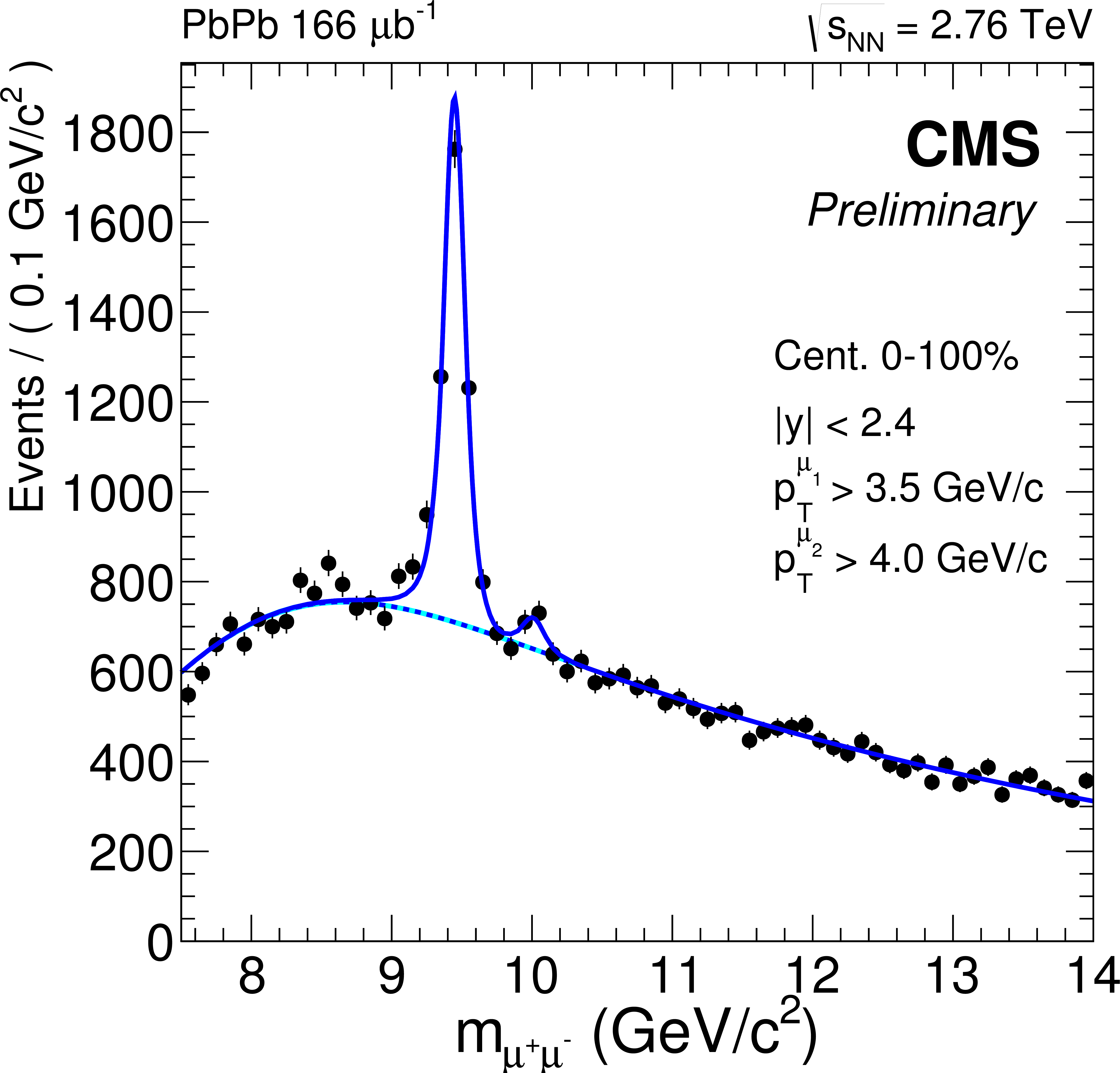}
    \includegraphics[width=0.33\textwidth]{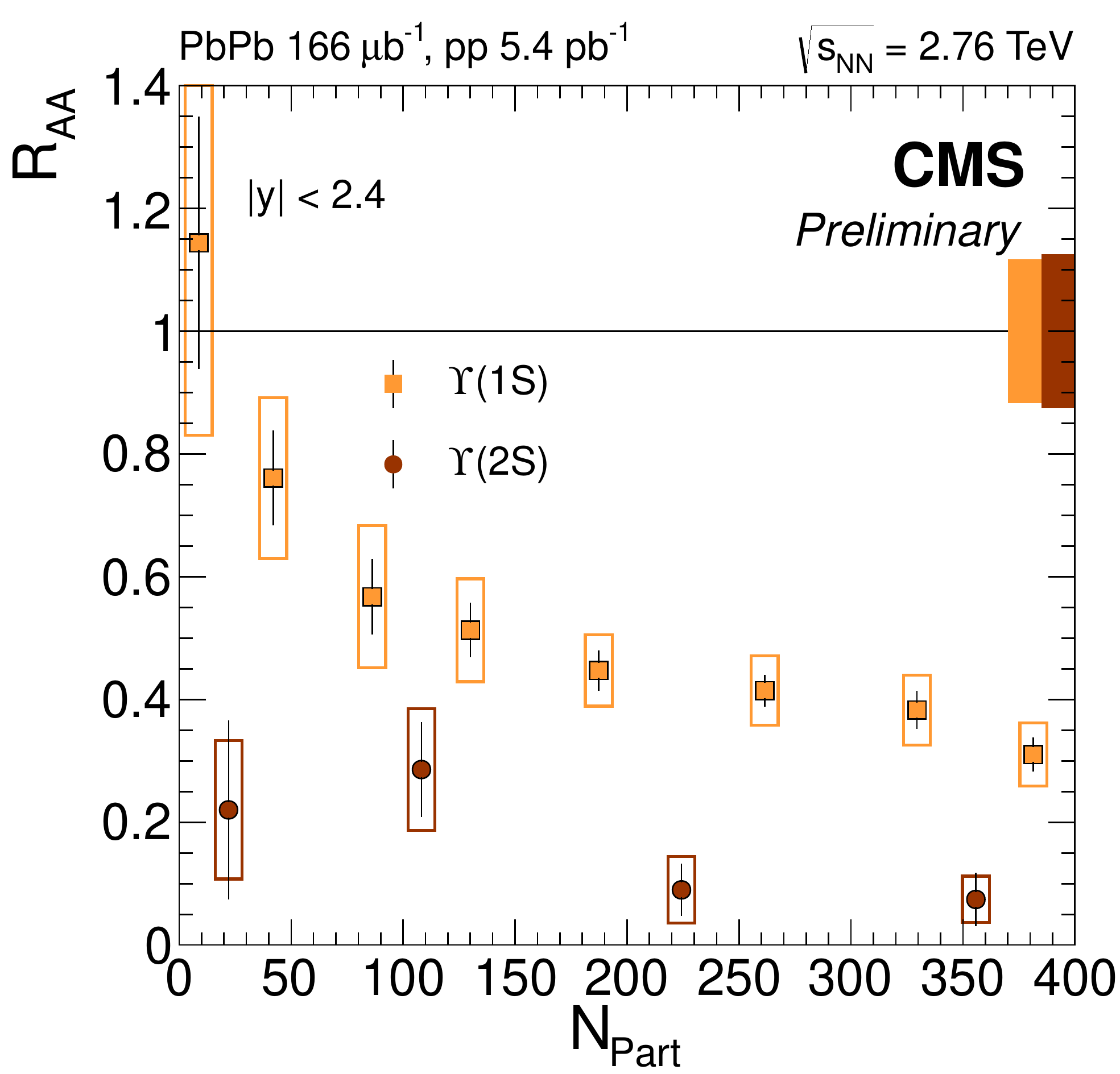}
    \caption{(Left) Dimuon invariant mass distributions for the pp and (Middle) \PbPb\  data at \sqrsNN\ 2.76 TeV.
    (Right) \RAA\ of $\Upsilon$(1S) and $\Upsilon$(2S) as a function of collision centrality.
    Figures from~\cite{CMS-PAS-HIN-15-001}.}
    \label{fig:upsilons}
\end{figure}

The first high-statistics results on $\Upsilon$ production in heavy-ion collisions were presented by CMS in \cite{Chatrchyan:2011pe}. 
Figure~\ref{fig:upsilons} shows the dimuon invariant-mass spectra in the $\Upsilon$ mass range for \PbPb\ (left) and pp (middle) \cite{CMS-PAS-HIN-15-001} 
at \sqrsNN\ 2.76 TeV collision energy for both systems.
It can be seen that the most tightly bound $\Upsilon(1S)$ state is clearly visible for both \PbPb\ and \pp\ systems (leftmost peak), whereas the $\Upsilon(2S)$ and  $\Upsilon(3S)$ states, observed in \pp\ collisions, are strongly suppressed in \PbPb. 
In Fig.~\ref{fig:upsilons}-right the suppression of $\Upsilon(1S)$ and $\Upsilon(2S)$ as a function of centrality quantified by the \RAA\ is presented.
The measurements confirm a sequential suppression of the observed bound states,  $\Upsilon(1S)$ (\RAA $=0.43 \pm 0.03 \pm 0.07$),  $\Upsilon(2S)$ (\RAA $=0.13 \pm 0.03 \pm 0.02$) and  $\Upsilon(3S)$ (with upper limit \RAA $=0.14$ at 95\% CL), as expected in the scenario of sequential melting \cite{CMS-PAS-HIN-15-001}. Further measurements show that feed-down from excited states (see \cite{Andronic:2015wma} for a review) seem not sufficient to explain the observed $\Upsilon(1S)$ \RAA, which may suggest a possible suppression also for this strongly-bound state, indicating the very high  temperature of the produced QGP, in the range 2--5$T_c$, as deduced from Fig.~\ref{fig:quarkonium_intro_plots}-left.

\paragraph{d. Heavy flavour in pp}

Heavy-flavour measurements in \pp\ collisions provide an important testing ground for various aspects of QCD. Because a hard scale is already introduced by their heavy mass, the partonic hard scattering processes can be calculated in the pQCD framework down to low \Pt\ and the total cross section can be computed integrating over \Pt. On the other hand, the formation of a quarkonium state is a non-perturbative process because it involves long distances and soft momentum scales. Hence, heavy-flavour studies can test both aspects, perturbative and non-perturbative, of QCD calculations.   

The total charm and beauty production cross sections, are found, within uncertainties, to be well described by pQCD calculations, as a function of the collision energy and including the LHC data \cite{Abelev:2012vra,Abelev:2014hla}.

Understanding heavy-flavour production in pp
collisions is crucial for the interpretation of the heavy-flavour measurements in AA collisions.
The correlation of open and hidden heavy-flavour yields to the charged-particle multiplicity can provide
important insight into the production mechanism of heavy quarks
and probe the interplay of hard and soft QCD processes that are responsible for particle production. They can provide information on the role of multi-parton interactions (MPI) where several hard parton-parton interactions take place in the same event with no correlations among them \cite{Alner:1985wj, Wang:1991us, Sjostrand:1987su, Bartalini:2010su}.
In addition, heavy-flavour production  could be affected from final-state effects because of the high-multiplicity environment of high-energy pp collisions \cite{Werner:2010ny, Lang:2013ex}. 

The yields of inclusive \Jpsi, non-prompt \Jpsi\ and prompt $D$ mesons were measured in \pp\ at $\sqrt{s} =$ 7 TeV as a function of the charged-particle multiplicity and were found to increase with increasing multiplicity for measurements at midrapidity. 
In particular, the average $D$ meson relative yields\footnote{A ratio between yields in a given multiplicity interval normalised to the multiplicity-integrated ones (relative yields).  The relative yields are reported as a function of the multiplicity of charged particles normalised to the average value for inelastic collisions (relative charged-particle multiplicity) \cite{Adam:2015ota}.} show a faster than linear increase at the highest multiplicities \cite{Adam:2015ota} and are quantitatively described by model calculations \cite{Ferreiro:2012fb} including contributions of MPI to particle production. The observed increase can also be interpreted in terms of the event hadronic activity 
which is accompanying the production of the heavy-flavour hadrons. 
The observed similarity of the enhancement of open and hidden heavy flavour yields suggests
that the enhancement is most likely related to the heavy-quark production mechanisms rather than hadronisation.
Further measurements, in progress, of \Jpsi\ production in different rapidity regions indicate the expected differences because of the different physics mechanisms in play at different rapidity domains.
Such studies have been extended with the measurement of cross sections of $\Upsilon$ normalized by their event activity integrated values in \PbPb\ collisions at \sqrsNN\ 2.76 TeV, which also show an increase with increasing charged-particle multiplicity \cite{Chatrchyan:2013nza}. 
In addition, analogous measurements are being performed for \pPb\ collisions where similar trends are also observed.
While different aspects of the analysis are under active investigation 
these unexpected intriguing results will have a significant impact on the understanding of the heavy-flavour production mechanism and the interpretation of the \pPb\ and \PbPb\ results.

\vspace{0.2cm}
In summary, the abundant production of heavy flavours at LHC opened up the possibility for detailed studies of QCD in high-energy hadronic collisions; from their production mechanisms in pp collisions, and their modification in pA collisions to 
nailing down the properties of the hot and dense strongly-interacting QGP matter in AA collisions.

\section{Summary}

With the Heavy-Ion Standard Model having passed its first tests 
a quantitative and systematic study of the sQGP has been (and is currently being)  carried out at LHC with the aim to measure with better precision its properties and parameters 
(like equation of state, viscosity, opacity).
For such precision measurements the LHC took full advantage of the 
huge increase in beam energy, by a factor of 10 relative to RHIC (with the associated higher particle density facilitating precision measurements of flow observables and the larger cross sections for hard probes) thanks to a powerful new generation of state-of-the-art experiments (ALICE, ATLAS, CMS, LHCb), characterized by excellent vertexing, tracking, particle identification, and also large phase-space coverage in \pt\ and rapidity.

In this article we have presented an overview of the results of hard probes from \PbPb\ collisions at the CERN LHC at collision energy of \sqrsNN\ 2.76 TeV as well as studies of the reference pp and \pPb\ systems. 
Jet quenching and heavy-quark measurements probe the QGP properties over a wide range of length scales and can therefore provide information not accessible via other measurements. 
An impressive range of novel results, some accessible for the first time, were obtained.
The first measurements of fully-reconstructed jets at the LHC contributed to the determination of the medium transport coefficient $\hat{q}$ with reduced systematic uncertainty. 
Measurements of heavy-flavour azimuthal anisotropies indicate that a significant fraction of the produced heavy
quarks  diffuse in the strongly-coupled QGP and are carried
along with it as it flows. 
The trend of the quarkonia suppression at small transverse momenta, 
observed from SPS to LHC energies,
established deconfinement in the quark-gluon plasma
and allows quantitative studies.
The study of charmonia enhanced the understanding of the interplay of different mechanisms
and  the role of recombination.
The measured sequential suppression of the bottomonium family demonstrated
the prediction of the dissociation pattern of the quarkonia states depending on their binding energy.  
The question of possible recombination and thermalization also of the $b$-quarks is being investigated.
Further progress of such studies will be achieved with a precision measurement of the total charm production cross section
by measuring the different open-charm decay channels down to the lowest possible \pT\ to minimize extrapolation errors. 
Last but not least,
the development of new observables for jet measurements and adapting well-tested tools, 
typically used to study  properties of jets in pp collisions, 
is expected to help establishing well-defined observables that are measurable and calculable 
with well-controlled precision,
and hence, a direct link between theory and experimental measurements.

To better present a
snapshot of the current state of the field, many preliminary measurements were included, while many  more interesting, new results are expected from analyses in progress.
Moreover, the LHC collaborations are in the process of upgrading the present detectors with the aim to enhance the precision of the measurements related to hard probes. In particular, improving vertexing and tracking capabilities is especially important for the detection of heavy flavoured observables and extending their kinematic converge at low \pt . In addition, developments are targeting data taking  at considerably higher rates. This will provide an  unprecedented amount of data, allowing even more precise measurements of hard probes, which will result in a yet better understanding of the QGP.

Overall, heavy-ion research, actively being pursued by an increasing number of scientists, at different research facilities is greatly contributing in understanding the nuclear aspects of nature.
Hard probes, referred to as ``rare probes" in the past, are presently abundantly produced at LHC, allowing us to perform multi-differential measurements, which have already provided significant insights into the physics of the QGP and they promise even more interesting results in the future.


\section*{Acknowledgements}
We thank Ralf Averbeck, Leticia Cunqueiro Mendez, 
Davide Caffarri, Thomas Peitzmann and Urs Wiedemann for useful discussions, and
Roberta Arnaldi, Alexander Milov, Adam Kisiel, {\L}ukasz Graczykowski, Camelia Mironov, Daniel Kiko{\l}a, Orlando Villalobos Baillie, Aleksi Vourinen for critical reading of the manuscript and most useful suggestions. 
The work of M. Janik was supported by the Polish National Science Centre under decisions no. 2013/08/M/ST2/00598, no. 2014/13/B/ST2/04054, and no. 2015/19/D/ST2/01600.

\clearpage \newpage
\section*{Appendix}

\begin{table*}[!hbt] 
\centering 
\caption{Reconstructed jet published results obtained in AA from LHC experiments. The experiment, the measurement, the colliding system, the energy in the centre-of-mass system $\snn$, R (the jet cone radius), the observables as well as the references are given.} 
\small
\label{Tab-jets} 
\begin{tabular*}{\textwidth}{|l|l|l|l|l|l|}
\hline 
Measurement & Colliding & \snn & R &  Observables  & Ref. \hspace{1.195cm} \\
& system & (Tev) & & (variables) & \\ 
\hline
\multicolumn{6}{c}{ALICE} \\

\hline 
Charged jets & \PbPb & 2.76 & 0.2, 0.3 & yields(\pt,cent.), $R_{CP}$(\pt), $R_{CP}(\rm cent.)$ &  \cite{Abelev:2013kqa}   \\ 
 & &  & 0.2 & $v_{2}^{\rm ch jet}$(\pt,cent.) & \cite{Adam:2015mda}\\
Charged + neutral jets & &  & 0.2 & yields(\pt,cent.), \RAA(\pt,cent.) &  \cite{Adam:2015ewa} \\
Hadron-jet &  &  & 0.2, 0.4, 0.5   &$\Delta_{\rm recoil}$, $\Delta I_{\rm AA}$ & \cite{Adam:2015doa}\\ 
\hline
\multicolumn{6}{c}{CMS}\\ 
\hline

Particle flow jets &  \PbPb & 2.76 & 0.2, 0.3, 0.4 & yields(\pt,cent.), \RAA(\pt,cent.)& \cite{Khachatryan:2016jfl}\\
Dijets & & & 0.5 & Ev. frac. (\pt$^{\rm leading\ jet}$, $\Delta\phi_{1,2}$, $A_J$),  & \cite{Chatrchyan:2011sx}\\
& & & & $\left < p^{||}_T \right >(A_J, \rm cent., \Delta R)$; $\left < (p_{T,1}-p_{T,2})/p_{T,1} \right >$(\pt$_{,1}$) & \\
&  &  & 0.3  & Ev. frac. ($\Delta\phi_{1,2}$, $A_J$, $x_j=p_{T,2}/p_{T,1}$) &  \cite{Chatrchyan:2012nia}  \\
& & & & $\left < p_{T,2}/p_{T,1} \right >$(\pt$_{,1}$) & \\
&  &  & 0.2--0.5 &  $\left <\not{p_{ T}^{||}} \right >(\Delta, A_J)$ & \cite{Khachatryan:2015lha} \\
Photon-jet &   &  & 0.3  & distribution of  $x_{J\gamma}=p_T^{\rm Jet}/p_T^{\gamma}$ &  \cite{Chatrchyan:2012gt} \\
Jet fragmentation  &  &  & 0.3  & fragm. fun. $\xi=ln(1/z)$ (\pt$>4$ GeV/$c$)& \cite{Chatrchyan:2012gw} \\
 &  &  & & fragm. fun. $\xi=ln(1/z)$ (\pt$>1$ GeV/$c$)& \cite{Chatrchyan:2014ava}  \\
Jet shapes &   &  & 0.3   & $\rho(r)$ & \cite{Chatrchyan:2013kwa} \\
Jet-track correlations  & & & 0.3 & jet-track correlations (\pt,$\Delta\eta,\Delta\phi$)  &\cite{Khachatryan:2016erx}\\
& & & 0.3 & redistribution of mom. in dijet events (\pt,$\Delta\phi$) & \cite{Khachatryan:2016tfj}\\

\hline
\multicolumn{6}{c}{ATLAS}\\ 
\hline
Inclusive jets & \PbPb & 2.76 & 0.2 & \RAA$^{\rm jet}$(\pt, $|y|$, cent.) &  \cite{Aad:2014bxa}   \\
&  &  &  & $v_2^{\rm jet}$(\pt, cent.) & \cite{Aad:2013sla}   \\
Dijets &   & & 0.4 & distribution of $A_J$,$\Delta\phi$ & \cite{Aad:2010bu}\\
Jet size &  &  & 0.2 - 0.5 & $R^R_{CP}/R^{0.2}_{CP}$(\pt) &  \cite{Aad:2012vca} \\
Jet fragmentation &  & & 0.4 & $D(z)$, $R_{D(z)}(z,p_T)$ & \cite{Aad:2014wha}  \\
Neighbouring jets & &  & 0.2, 0.3, 0.4 & ${\rm d} R_{\Delta R}/{\rm d}E_{T}^{\rm nbr}(E_{T}^{\rm nbr},{\rm cent.})$, $\rho_{R_{\Delta R}}(E_{T}^{\rm nbr},{\rm cent.})$ &  \cite{Aad:2015bsa}  \\
\hline

\end{tabular*}
\end{table*}

\begin{table*}[!t] 
\centering
 \caption{Open heavy flavour  published results obtained in AA from LHC experiments. The probe, the energy in the centre-of-mass system (\snn), the covered kinematic ranges, the observables and references are given.
 The tables are taken from  \cite{Andronic:2015wma} 
and were updated with new results.} 
\small
 \label{tab:OpenHeavy_expSummary_LHC}
\begin{tabular*}{\textwidth}{|l|l|l|l|l|l|l|}
 \hline 
 Probe & Colliding & \snn &  $y_{\rm cms}$ (or $\eta_{\rm cms}$) & \pt & Observables & Ref. \hspace{0.85cm} \\ 
 & system & (\TeV) &  & (\GeVc) & & \\ 
 \hline 
 \multicolumn{7}{c}{ALICE} \\ 
 \hline 
\Dzero, \Dplus, \Dstarplus & \pb & 2.76 & $|y|<0.5$ & 2 -- 16 & yields\,(\pt), $\raa$(\pt) & \cite{ALICE:2012ab}\\ 
 & & & & 2 -- 12 & $\raa({\rm centrality})$ & \\  
 & & & & 6 -- 12 & $\raa({\rm centrality})$ & \\ 
 & & & $|y|<0.8$ & 2 -- 16 &  \vtwo(\pt), \vtwo(centrality,\pt), $\raa^{\text{in/out plane}}$(\pt) & \cite{Abelev:2014ipa,Abelev:2013lca}\\  
  & & & $|y|<0.5$ & 5 -- 8 &  \RAA(centrality) &  \cite{Adam:2015nna}\\ 
  & & &           & 8 -- 16 &  \RAA(centrality) & \\ 
  & & &           & 1 -- 36 &  yields(\pt) & \cite{Adam:2015sza}\\ 
  & & &           & 1 -- 20 &  \RAA(\pt) & \\ 
$D_{s}^{+}$ &  &  & $|y|<0.5$ & 4 -- 12 & yields\,(\pt) & \cite{Adam:2015jda}\\ 
  & & &           & 4 -- 12 &  \RAA(\pt) & \\ 
 \hfm &  &  & $2.5<y<4$ & 4 -- 10 & $\raa$(\pt) & \cite{Abelev:2012qh}\\  
 & & & & 6 -- 10 &  $\raa({\rm centrality})$ & \\ 
  & & & & 3 -- 10 &  \vtwo$({\rm centrality})$ & \cite{Adam:2015pga}\\  
 & & & &  3 -- 10 &  \vtwo(\pt)  & \\  
 \hfe &  &  & $|y|<0.7$ & 0.5 -- 13 & \vtwo(\pt), \vtwo(centrality) & \cite{Adam:2016ssk}\\
  &  &  & $|y|<0.6$ & 3 -- 18 & \RAA(\pt) & \cite{Adam:2016khe}\\
 \ensuremath{{\rm b}\ (\to c) \to e^{\pm}} & & &  $|y|<0.8$ & 1.3 -- 8 &  yields\,(\pt), \RAA(\pt) & \cite{Adam:2016wyz}\\
 non-prompt \jpsi &  &  & $|y|<0.8$ & 1.5 -- 10 & \raa(\pt) & \cite{Adam:2015rba} \\ 
 \hline 
 \multicolumn{7}{c}{CMS} \\ 
 \hline 
 $b$-jets & \pb & 2.76 & $|\eta|<2$ & 80 -- 250 & yields~(\pt) & \cite{Chatrchyan:2013exa}\\ 
 & & & & & $\raa$(\pt) & \\ 
 & & & & 80 -- 110 & $\raa({\rm centrality})$ & \\ 
 non-prompt \jpsi &  &  & $|y|<2.4$ & 6.5 -- 30 & yields~({\rm centrality}) & \cite{Chatrchyan:2012np}\\ 
 & & & & & $\raa$({\rm centrality}) & \\  
 & & & $|y|<2.4$ & 6.5 -- 30  & $\raa$, \vtwo$(\text{cent.},\,$\pt$,\,y)$ &\cite{Khachatryan:2016ypw}\\
      & & & $1.6<|y|<2.4$ & 3 -- 6.5 & & \\ 
 \hline
 \end{tabular*} 
\end{table*} 

\begin{table*}[!t] 
  \centering 
  \caption{Quarkonium results obtained in AA from LHC experiments. The 
    experiment, the probes, the collision energy, the covered kinematic range 
    and the observables are given.
    The tables are taken from  \cite{Andronic:2015wma} 
and were updated with new results.}
    \small
  \label{tab:Quarkonium_expSummary_LHC} 
  \begin{tabular*}{\textwidth}{|l|l|l|l|l|l|l|} 
    \hline 
    Probe & Colliding  & $\sqrt{s_{NN}}$  & $y$ & \pt  & Observables & Ref. \hspace{2.065cm} \\ 
      & system &  (TeV) &  & (GeV/$c$)& & \\ 
    \hline 
    \multicolumn{7}{c}{ALICE} \\ 
    \hline 
    \jpsi & \pb & 2.76 & $|y|<0.9$ & \pt$>0$ & $\raa(\text{cent.,\,}$\pt) & \cite{Abelev:2013ila,Adam:2015rba}\\  
    & & & $2.5<y<4$ & \pt$>0$ & \RAA$(\text{cent.,\,}$\pt$\text{,\,y)}$ & \cite{Abelev:2012rv,Abelev:2013ila,Adam:2015isa} \\  
    & & & & $0<$\pt$<10$ & \vtwo$(\text{cent.},\,$\pt) & \cite{ALICE:2013xna} \\ 
    
    \jpsi & & 5.02 & $2.5<y<4$ & \pt$>0$ & yield, $\raa(\text{cent.,\,}$\pt) &  \cite{Adam:2016rdg}\\
    
    \psiP & & 2.76 & & \pt$<3$ & $\frac{(N_{\psiP}/N_{J/\Psi})_{\mathrm{Pb-Pb}}}{(N_{\psiP}/N_{J/\Psi})_{\mathrm{pp}}}(\text{cent.})$ &  \cite{Adam:2015isa} \\ 
    & & & & $3<$\pt$<8$ & & \\ 

    \upsa & & & & \pt$>0$ & $\raa(\text{cent.},\,y)$ & \cite{Abelev:2014nua}\\ 
    \hline 
    \multicolumn{7}{c}{ATLAS} \\ 
    \hline 
    \jpsi & \pb & 2.76 & $|\eta|<2.5$ & \pt$\gtrsim6.5$ & $\rcp(\text{cent.})$ & \cite{Aad:2010aa}\\ 
    \hline 
    \multicolumn{7}{c}{CMS} \\ 
    \hline 
   
    \jpsi\ (prompt) & \pb & 2.76   & $|y|<2.4$ & $6.5<$\pt$<30$& yield, $\raa(\text{cent.},\,$\pt$,\,y)$ & \cite{Chatrchyan:2012np} \\  
    & & & & & \vtwo$(\text{cent.},\,$\pt$,\,y)$ & \cite{CMS:2013dla} \\ 
    & & & $1.6<|y|<2.4$ & $3<$\pt$<30$ & & \\ 
    & & & $|y|<1.2$ & $6.5<$\pt$<30$ & yield and \raa & \cite{Chatrchyan:2012np} \\ 
    & & & $1.2<|y|<1.6$ & $5.5<$\pt$<30$ & & \\ 
    & & & $1.6<|y|<2.4$ & $3<$\pt$<30$ & &\\ 
 
    & & & $|y|<2.4$ & $3 < $\pT $< 30$  & $\raa$, \vtwo$(\text{cent.},\,$\pt$,\,y)$ &\cite{Khachatryan:2016ypw}\\
      & & & $1.6<|y|<2.4$ & $3<$\pt$<30$ & & \\ 
    \jpsi,  \psiP & & 5.02 & $|y|<1.6$ & $6.5 < $\pT $< 30$ &  $\frac{(N_{\psiP}/N_{J/\Psi})_{\mathrm{Pb-Pb}}}{(N_{\psiP}/N_{J/\Psi})_{\mathrm{pp}}}(\text{cent.})$ & \cite{Sirunyan:2016znt}\\   
 (prompt)    & & & $1.6<|y|<2.4$ & $3<$\pT$<30$ & & \\
    \psiP (prompt) & & 2.76 & $|y|<1.6$ & $6.5<$\pt$<30$  & \raa, $\frac{(N_{\psiP}/N_{J/\Psi})_{\mathrm{Pb-Pb}}}{(N_{\psiP}/N_{J/\Psi})_{\mathrm{pp}}}(\text{cent.})$& \cite{Khachatryan:2014bva} \\ 
    & & &$1.6<|y|<2.4$ & $3<$\pt$<30$ & & \\ 
    \upsa & & & $|y|<2.4$ & \pt$>0$ & yield, $\raa(\text{cent.},\,$\pt$,\,y)$ & \cite{Chatrchyan:2012np} \\ 
    \upsn & & & $|y|<2.4$ & \pt$>0$ & $\raa(\text{cent.})$ & \cite{Chatrchyan:2011pe,Chatrchyan:2012lxa}  \\  
    & & & & & $\frac{(N_{\upsb}/N_{\upsa})_{\mathrm{Pb-Pb}}}{(N_{\upsb}/N_{\upsa})_{\mathrm{pp}}}(\text{cent.})$ & \cite{Chatrchyan:2013nza}  \\  
     & & &  & \pt$< 20$ &  yield, $\raa(\text{cent.},\,$\pt$,\,y)$ & \cite{Khachatryan:2016xxp}  \\

    \hline 
  \end{tabular*} 
\end{table*}

\clearpage \newpage
\section*{References}
\bibliographystyle{elsarticle-num}
\bibliography{bibliografia}

\end{document}